\documentclass[12pt, reqno]{amsart}

\usepackage[margin=1.in, a4paper, foot=.3in]{geometry}

\usepackage{graphicx, color}

\definecolor{darkred}{rgb}{.5,0,0}
\definecolor{darkgreen}{rgb}{0,0.5,0}
\definecolor{darkblue}{rgb}{0,0,.5}

\usepackage{graphicx, color}

\usepackage{hyperref}
\hypersetup {
    pdfstartview=FitH,
    colorlinks,
    citecolor=darkgreen,
    linkcolor=darkred,
    urlcolor=darkblue
}

\usepackage{cite}

\usepackage[noBBpl,slantedGreek]{mathpazo}
\usepackage[mathcal]{euscript}
\usepackage{bm}

\usepackage{amssymb}
\usepackage{amsmath}
\usepackage{stmaryrd}
\usepackage{mathtools}

\allowdisplaybreaks

\numberwithin{equation}{section}

\newcommand {\bbC}{\mathbb C}
\newcommand {\bbE}{\mathbb E}
\newcommand {\bbF}{\mathbb F}
\newcommand {\bbN}{\mathbb N}
\newcommand {\bbO}{\mathbb O}
\newcommand {\bbZ}{\mathbb Z}

\newcommand {\calL}{\mathcal L}
\newcommand {\calO}{\mathcal O}

\newcommand {\gothg}{\mathfrak g}
\newcommand {\gothgl}{\mathfrak g}
\newcommand {\gothh}{\mathfrak h}
\newcommand {\gothk}{\mathfrak k}
\newcommand {\gothsl}{\mathfrak{sl}}
\newcommand {\hgothh}{\widehat{\mathfrak h}}
\newcommand {\tgothh}{\widetilde{\mathfrak h}}

\newcommand {\gliii}{\mathfrak{gl}_3}
\newcommand {\slii}{\mathfrak{sl}_2}
\newcommand {\sliii}{\mathfrak{sl}_3}
\newcommand {\uqbm}{\mathrm U_q(\mathfrak b^-)}
\newcommand {\uqbp}{\mathrm U_q(\mathfrak b^+)}
\newcommand {\uqhg}{\mathrm U_q(\widehat{\mathfrak g})}
\newcommand {\uqg}{\mathrm U_q(\mathfrak g)}
\newcommand {\uqglii}{\mathrm U_q(\mathfrak{gl}_2)}
\newcommand {\uqgliii}{\mathrm U_q(\mathfrak{gl}_3)}
\newcommand {\uqgllpo}{\mathrm U_q(\mathfrak{gl}_{l + 1})}
\newcommand {\uqlg}{\mathrm U_q(\mathcal L(\mathfrak g))}
\newcommand {\uqlslii}{\mathrm U_q(\mathcal L(\mathfrak{sl}_2))}

\newcommand {\uqlsllpo}{\mathrm U_q(\mathcal L(\mathfrak{sl}_{l + 1}))}
\newcommand {\uqtg}{\mathrm U_q(\widetilde{\mathfrak g})}

\newcommand {\mbar}[3]{\hskip #2 \overline{\hskip -#2 #1 \hskip -#3} \hskip #3}

\newcommand {\obmPsi}{\mbar{\bm \Psi}{.1em}{.01em}}
\newcommand {\oPsi}{\mbar{\Psi}{.1em}{.01em}}
\newcommand {\otheta}{\mbar{\theta}{.1em}{.01em}}
\newcommand {\oV}{\mbar{V}{.1em}{.01em}}
\newcommand {\ovarphi}{\mbar{\varphi}{0.em}{.02em}}

\newcommand {\qbinom}[2]{\begin{bmatrix} #1 \\ #2 \end{bmatrix}}

\DeclareMathOperator {\diag}{diag}
\DeclareMathOperator {\hght}{ht}
\DeclareMathOperator {\Osc}{Osc}

\title{Oscillator versus prefundamental representations}

\author[H. Boos]{Hermann Boos}
\address{Fachbereich C -- Physik, Bergische Universit\"at Wuppertal, 42097 Wuppertal, Germany}
\email{hboos@uni-wuppertal.de}

\author[F. G\"ohmann]{Frank G\"ohmann}
\address{Fachbereich C -- Physik, Bergische Universit\"at Wuppertal, 42097 Wuppertal, Germany}
\email{goehmann@uni-wuppertal.de}

\author[A. Kl\"umper]{Andreas Kl\"umper}
\address{Fachbereich C -- Physik, Bergische Universit\"at Wuppertal, 42097 Wuppertal, Germany}
\email{kluemper@uni-wuppertal.de}

\author[Kh. S. Nirov]{\vskip .2em Khazret S. Nirov}
\address{Institute for Nuclear Research of the Russian Academy of Sciences, 60th October Ave 7a, 117312 Moscow, Russia}
\curraddr{Fachbereich C -- Physik, Bergische Universit\"at Wuppertal, 42097 Wuppertal, Germany}
\email{nirov@uni-wuppertal.de}

\author[A. V. Razumov]{Alexander V. Razumov}
\address{Institute for High Energy Physics, NRC "Kurchatov Institute", 142281 Protvino, Mos\-cow region, Russia}
\email{Alexander.Razumov@ihep.ru}

\begin{document}

\addtolength {\jot}{3pt}

\begin{abstract}
For the case of quantum loop algebras $\mathrm U_q(\mathcal L(\mathfrak{sl}_{l + 1}))$ with $l = 1, 2$ we find the $\ell$-weights and the corresponding $\ell$-weight vectors for the representations obtained via Jimbo's homomorphism, known also as evaluation representations. Then we find the $\ell$-weights and the $\ell$-weight vectors for the $q$-oscillator representations of Borel subalgebras of the same quantum loop algebras. This allows, in particular, to relate $q$-oscillator and prefundamental representations.
\end{abstract}

\maketitle

\tableofcontents

\section{Introduction}

One of the modern methods to investigate quantum integrable system is based on the notion of a quantum group. To be more exact, one should say that here a special class of quantum groups, called quantum loop algebras, is used, see section \ref{s:qla} for the definition. For the first time the method was consistently used to construct integrability objects, such as monodromy operators and $L$-operators, and for the proof of functional relations by Bazhanov, Lukyanov and Zamolodchikov \cite{BazLukZam96, BazLukZam97, BazLukZam99}. They investigated the quantum version of KdV theory. Later the method proved to be efficient for studying various quantum integrable models. With its help one constructs $R$-operators \cite{KhoTol92, LevSoiStu93, ZhaGou94, BraGouZhaDel94, BraGouZha95, BooGoeKluNirRaz10, BooGoeKluNirRaz11}, monodromy operators and $L$-operators \cite{BazTsu08, BooGoeKluNirRaz10, BooGoeKluNirRaz11, BooGoeKluNirRaz13, Raz13, BooGoeKluNirRaz14a}, and proves functional relations \cite{BazHibKho02, Koj08, BazTsu08, BooGoeKluNirRaz14a, BooGoeKluNirRaz14b, NirRaz14}.

The central object of the approach is the universal $R$-matrix being an element of the tensor product of two copies of the quantum loop algebra. The integrability objects are constructed by the choice of representations for the factors of that tensor product. In fact, the universal $R$-matrix is an element of the tensor product of two different Borel subalgebras of the quantum group. Certainly, representations of the Borel subalgebras can be constructed by restricting representations of the full algebra. Such representations are used to define various monodromy operators. However, one needs more representations. For example, to construct $L$-operators one uses the so called $q$-oscillator representations which can be obtained from the representations used to construct the monodromy operators via some limiting procedure \cite{BazHibKho02, BooGoeKluNirRaz13, BooGoeKluNirRaz14a, BooGoeKluNirRaz14b, NirRaz14}.

Recently, Hernandez and Jimbo constructed some representations of the Borel subalgebras of quantum loop algebras as a limit of the Kirillov-Reshetikhin modules \cite{HerJim12}. It is common now to call these representations prefundamental \cite{FreHer15}.

For the study of representations of quantum loop algebras and their Borel subalgebras the notion of $\ell$-weights and $\ell$-weight vectors appear very useful \cite{HerJim12, FreHer15, MukYou14}. In the present paper we find the $\ell$-weights and the corresponding $\ell$-weight vectors for representations of quantum loop algebras $\uqlsllpo$ with $l = 1, 2$ obtained via Jimbo's homomorphism, known also as evaluation representations. Then we find the $\ell$-weights and the $\ell$-weight vectors for the $q$-oscillator representations of Borel subalgebras of the same quantum loop algebras. This allows, in particular, to relate the $q$-oscillator and prefundamental representations. In addition, we demonstrate how the knowledge of the $\ell$-weights allows one to relate the $q$-oscillator representations with the evaluation representations. This is important for the investigation of the corresponding quantum integrable systems.

The definition of a quantum loop algebra via Drinfeld-Jimbo generators is symmetric with respect to the change of the deformation parameter $q$ to $q^{-1}$. This is not so for the definition of the coproduct and antipode. The second Drinfeld's realization used to define $\ell$-weights and $\ell$-weight vectors is not symmetric with respect to this change as well. In the present paper we follow the definitions usually used in the papers on representations of quantum loop algebras. In fact we used the opposite conventions in our previous papers on applications of quantum groups to investigation of quantum integrable systems. Therefore, when we need formulas from our previous papers we first change $q$ to $q^{-1}$ and then use them.

\section{Quantum loop algebras} \label{s:qla}

\subsection{Drinfeld--Jimbo definition}

Let $A = (a_{i j})_{i, \, j = 1}^l$ be the generalized Cartan matrix of finite type and $\widehat A = (a_{i j})_{i, \, j = 0}^l$ the corresponding generalized Cartan matrix of untwisted affine type. We denote by $\gothg$ and $\widehat \gothg$ the corresponding Kac--Moody algebras and use the natural identification of $\gothg$ with a subalgebra of $\widehat \gothg$.

We denote by $\calL(\gothg)$ the loop algebra of $\gothg$, and by $\widetilde \gothg$ its standard central extension by a one-dimensional centre $\bbC \, c$.  It can be shown that the Lie algebra $\widehat \gothg$ is isomorphic to the Lie algebra obtained from $\widetilde \gothg$ by adding a natural derivation $d$. We will identify these Lie algebras \cite{Kac90}.

We define $I = \{1, \, \ldots, l\}$ and $\widehat I = \{0, \, 1, \ldots, l\}$, so that $A = (a_{i j})_{i, \, j \in I}$ and $\widehat A = (a_{i j})_{i, \, j \in \widehat I}$. Denote by $D$ the unique diagonal matrix $\diag(d_0, \, d_1, \, \ldots, d_l)$ such that the matrix $B = (b_{i j})_{i, j \in \widehat I} = D A$ is symmetric and $d_i$, $i \in \widehat I$, are relatively prime positive integers.
 
 Let $h_i$, $i \in I$, be the Cartan generators of $\gothg$, and $h_i$, $i \in \widehat I$, $d$ the Cartan generators of $\widehat \gothg$. Hence, the Cartan subalgebras of $\gothg$ and $\widehat \gothg$ are
\begin{equation*}
\gothh = \bigoplus_{i \in I} \bbC \, h_i, \qquad \widehat \gothh = \Bigl( \bigoplus_{i \in \widehat I} \bbC \, h_i \Bigr) \oplus \bbC \, d.
\end{equation*}
In fact, we have
\begin{equation*}
\widehat \gothh = \gothh \oplus \bbC \, c \oplus \bbC \, d.
\end{equation*}
We identify the space $\gothh^*$ with the subspace of $\widehat \gothh^*$ defined as
\begin{equation*}
\gothh^* = \{ \gamma \in \widehat \gothh^* \mid \langle \gamma, \, c \rangle = 0, \quad \langle \gamma, \, d \rangle = 0 \}.
\end{equation*}
It is also convenient to denote
\begin{equation*}
\tgothh = \gothh \oplus \bbC \, c = \Bigl( \bigoplus_{i \in I} \bbC \, h_i \Bigr) \oplus \bbC \, c = \bigoplus_{i \in \widehat I} \bbC \, h_i.
\end{equation*}
and identify the space $\gothh^*$ with the subspace of $\widetilde \gothh^*$ which consists of the elements $\gamma \in \widetilde \gothh^*$ satisfying the condition
\begin{equation}
\langle \gamma, \, c \rangle = 0. \label{lambdac}
\end{equation}

The simple roots $\alpha_i \in \gothh^*$, $i \in I$, of $\gothg$ are given by the relations
\begin{equation*}
\langle \alpha_i, \, h_j \rangle = a_{j i}, \quad i, \, j \in I,
\end{equation*}
while for the simple roots $\alpha_i \in \widehat \gothh^*$, $i \in \widehat I$, of $\widehat \gothg$ we have the definition
\begin{equation*}
\langle \alpha_i, \, h_j \rangle = a_{j i}, \quad i, j \in \widehat I, \qquad \langle \alpha_0, \, d \rangle = 1, \qquad \langle \alpha_i, \, d \rangle = 0, \quad i \in I.
\end{equation*}
We fix the non-degenerate symmetric bilinear form on $\gothh$ by the relations
\begin{equation*}
(h_i | h_j) = a^{\mathstrut}_{i j} \, d_j^{-1},
\end{equation*}
and on $\widehat \gothh$ by
\begin{equation*}
(h_i | h_j) = a^{\mathstrut}_{i j} \, d_j^{-1}, \qquad (h_i | d) = \delta^{\mathstrut}_{i0} \, d_0^{-1}, \qquad (d | d) = 0.
\end{equation*}

Let $\hbar$ be a nonzero complex number such that $q = \exp \hbar$ is not a root of unity. For each $i \in \widehat I$ we set
\begin{equation*}
q_i = q^{d_i}. 
\end{equation*}
The {\em quantum group\/} $\uqhg$ is a unital associative $\bbC$-algebra generated by the elements $e_i$, $f_i$, $i \in \widehat I$, and $q^x$, $x \in \hgothh$, with the relations
\begin{gather}
q^0 = 1, \qquad q^{x_1} q^{x_2} = q^{x_1 + x_2}, \label{djra} \\
q^x e_i \, q^{-x} = q^{\langle \alpha_i, \, x \rangle} e_i, \qquad q^x f_i \, q^{-x} = q^{- \langle \alpha_i, \, x \rangle} f_i, \\
[e_i, \, f_j] = \delta_{i j} \, \frac{q_i^{h_i} - q_i^{- h_i}}{q^{\mathstrut}_i - q_i^{-1}}, \\
\sum_{n = 0}^{1 - a_{i j}} (-1)^n e_i^{(1 - a_{i j} - n)} e^{\mathstrut}_j \, e_i^{(n)} = 0, \qquad \sum_{n = 0}^{1 - a_{i j}} (-1)^n f_i^{(1 - a_{i j} - n)} f^{\mathstrut}_j \, f_i^{(n)} = 0, \label{djrd}
\end{gather}
where $e_i^{(n)} = e_i^n / [n]_{q_i}!$, $f_i^{(n)} = f_i^n / [n]_{q_i}!$. Here and below we use the standard notations for $q$-numbers
\begin{gather*}
[\nu]_q = \frac{q^\nu - q^{-\nu}}{q - q^{-1}}, \quad \nu \in \bbC, \qquad [n]_q! = \prod_{k = 1}^n [k]_q, \quad n \in \bbZ_+, \\ \qbinom{n}{m}_q = \frac{[n]_q!}{[m]_q! [n - m]_q!}, \quad n, m \in \bbZ_+,
\end{gather*}
and assume that for any $\nu \in \bbC$
\begin{equation*}
q^\nu = \exp (\hbar \nu).
\end{equation*}
We will also use the notation
\[
\kappa_q = q - q^{-1}.
\]
The quantum group $\uqhg$ is a Hopf algebra with the comultiplication $\Delta$, the antipode $S$, and the counit $\varepsilon$ defined by the relations
\begin{gather}
\Delta(q^x) = q^x \otimes q^x, \qquad \Delta(e_i) = e_i \otimes 1 + q_i^{h_i} \otimes e_i, \qquad \Delta(f_i) = f_i \otimes q_i^{-h_i} + 1 \otimes f_i, \label{hsa} \\
S(q^x) = q^{- x}, \qquad S(e_i) = - q_i^{-h_i} e_i, \qquad S(f_i) = - f_i \, q_i^{h_i}, \\
\varepsilon(q^x) = 1, \qquad \varepsilon(e_i) = 0, \qquad \varepsilon(f_i) = 0. \label{hsc}
\end{gather}
We define the quantum group $\uqg$ as a Hopf subalgebra of $\uqhg$ generated by $e_i$, $f_i$, $i \in I$, and $q^x$, $x \in \gothh$. 

The quantum group $\uqhg$ has no nontrivial finite dimensional representations, and therefore we proceed to the consideration of the corresponding quantum loop algebra. As the first step we define the quantum group $\uqtg$ as a unital associative $\bbC$-algebra generated by the elements $e_i$, $f_i$, $i \in I$, and $q^x$, $x \in \tgothh$, with relations (\ref{djra})--(\ref{djrd}). Then, the {\em quantum loop algebra\/} $\uqlg$ is defined as the quotient algebra of $\uqtg$ by the two-sided Hopf ideal generated by the elements of the form $q^{\nu c} - 1$ with $\nu \in \bbC^\times$. It is convenient to consider the quantum group $\uqlg$ as a unital associative $\bbC$-algebra generated by the same generators as $\uqtg$ with relations (\ref{djra})--(\ref{djrd}) and additional relations
\begin{equation}
q^{\nu c} = 1, \qquad \nu \in \bbC^\times. \label{qnuc}
\end{equation}
The structure of a Hopf algebra on $\uqlg$ is again given by relations (\ref{hsa})--(\ref{hsc}).

\subsection{Cartan--Weyl generators}

Denote by $\triangle$ and $\widehat \triangle$ the root systems of $\gothg$ and $\widehat \gothg$ respectively. They are related in the following way \cite{Kac90}
\begin{equation*}
\widehat \triangle = \{\gamma + n \delta \mid \gamma \in \triangle, \ n \in \bbZ \} \cup \{n \delta \mid n \in \bbZ \setminus \{0\}\},
\end{equation*}
where $\delta = \alpha_0 + \theta$ with $\theta$ being the highest root of $\gothg$. The systems $\triangle_+$ and $\widehat \triangle_+$ of positive roots of $\gothg$ and $\widehat \gothg$ are related as
\begin{multline*}
\widehat \triangle_+ = \{\gamma + n \delta \mid  \gamma \in \triangle_+, \ n \in \bbZ_+\} \\ \cup \{n \delta \mid n \in \bbN\} \cup \{(\delta - \gamma) + n \delta \mid  \gamma \in \triangle_+, \ n \in \bbZ_+\}.
\end{multline*}
As usually, for the systems $\triangle_-$ and $\widehat \triangle_-$ of positive roots of $\gothg$ and $\widehat \gothg$ we have $\triangle_- = - \triangle_+$ and $\widehat \triangle_- = - \widehat \triangle_+$.

The abelian group
\begin{equation*}
\widehat Q = \bigoplus_{i \in \widehat I} \bbZ \, \alpha_i
\end{equation*}
is called the root lattice of $\widehat \gothg$. We also define
\begin{equation*}
\widehat Q_+ = \bigoplus_{i \in \widehat I} \bbZ_+ \, \alpha_i, \qquad \widehat Q_- = \bigoplus_{i \in \widehat I} \bbZ_- \, \alpha_i.
\end{equation*}

The algebra $\uqlg$ can be considered as $\widehat Q$-graded if we assume that
\begin{equation*}
e_i \in \uqlg_{\alpha_i}, \qquad f_i \in \uqlg_{- \alpha_i}, \qquad q^x \in \uqlg_0
\end{equation*}
for any $i \in \widehat I$ and $x \in \tgothh$. An element $a$ of $\uqlg$ is called a {\em root vector corresponding to a root\/} $\gamma$ of $\widehat \gothg$ if $a \in \uqlg_\gamma$. It is clear that $e_i$ and $f_i$ are root vectors corresponding to the roots $\alpha_i$ and $- \alpha_i$. One can find linearly independent root vectors corresponding to all roots of $\widehat \gothh$. These vectors, together with the elements $q^x$, $x \in \tgothh$, are called {\em Cartan--Weyl\/} generators of $\uqlg$. It appears that the ordered monomials constructed from the Cartan--Weyl generators form a Poincar\'e--Birkhoff--Witt basis of $\uqlg$.

We denote the Cartan--Weyl generator corresponding to a root $\gamma \in \widehat \triangle_+$ by $e_\gamma$, and the Cartan--Weyl generator corresponding to a root $\gamma \in \widehat \triangle_-$ by $f_{-\gamma}$. We assume that
\begin{equation*}
e_{\alpha_i} = e_i, \qquad f_{\alpha_i} = f_i.
\end{equation*}
It is convenient to write $e_{\delta - \theta}$ and $f_{\delta - \theta}$ instead of $e_{\alpha_0}$ and $f_{\alpha_0}$.

To define Cartan--Weyl generators corresponding to the remaining roots we use the me\-thod of Khoroshkin and Tolstoy \cite{TolKho92, KhoTol93}. For another approach we refer the reader to the paper~\cite{Bec94a}. 

First fix some normal order \cite{AshSmiTol79, Tol89} for $\widehat \triangle_+$ satisfying the conditions that the roots $n \delta$ are ordered in arbitrary way and that
\begin{equation*}
\gamma + n \delta \prec m \delta \prec (\delta - \gamma) + k \delta
\end{equation*}
for any $\gamma \in \triangle_+$ and $n, m, k \in \bbZ_+$.

Now introduce the notion of $q$-commutator $[ \, \ , \ ]_q$. Let $\alpha, \beta \in \widehat Q_+$, $a \in \uqlg_\alpha$ and $b \in \uqlg_\beta$. Define the $q$-commutator of $a$ and $b$ as\footnote{Remind that in comparison with our previous papers we change $q$ to $q^{-1}$.}
\begin{equation*}
[a, \, b]_q = a b - q^{- (\alpha | \beta)} b a.
\end{equation*}
For $\alpha, \beta \in \widehat Q_-$, $a \in \uqlg_\alpha$ and $b \in \uqlg_\beta$ we assume that
\begin{equation*}
[a, \, b]_q = a b - q^{(\alpha | \beta)} b a.
\end{equation*}

In general, the root vectors corresponding to the roots $\gamma \in \widehat \triangle_+$ and $- \gamma \in \widehat \triangle_-$ are defined as follows. Assume that $\gamma = \alpha + \beta$, $\alpha \prec \gamma \prec \beta$, and there are no other roots $\alpha' \succ \alpha$ and $\beta' \prec \beta$ such that $\gamma = \alpha' + \beta'$. If the root vectors $e_\alpha$, $e_\beta$ and $f_\alpha$, $f_\beta$ are already defined, then we put
\begin{equation*}
e_\gamma = [e_\alpha, \, e_\beta]_q, \qquad f_\gamma = [f_\beta, \, f_\alpha]_q .
\label{e1}
\end{equation*}

To define the root vectors corresponding to the roots from $\triangle$ we use the following iterative procedure.  Recall that the {\em height of a root\/} $\gamma = \sum_{i \in I} k_i \alpha_i \in \Delta_+$ is defined as
\begin{equation*}
\hght \gamma = \sum_i k_i.
\end{equation*}
Note that $\theta$ is a unique positive root of the highest height. Assume that for some number $m$, such that $1 \le m < \hght \theta$, the root vectors $e_\gamma$ and $f_\gamma$ for all $\gamma \in \triangle_+$ with $1 \le \hght \gamma \le m$ are already defined. Let $\gamma \in \triangle_+$ and $\hght \gamma = m + 1$. It can be shown that for some $i \in I$ the root $\gamma$ can be represented as
\begin{equation*}
\gamma = \alpha_i + \beta
\end{equation*}
where $\beta \in \triangle_+$ and $\hght \beta = m$. Fixing such a representation, we define
\begin{align*}
& e_\gamma = \begin{cases}
[e_{\alpha_i}, \, e_\beta]_q & \alpha_i \prec \beta \\
[e_\beta, \, e_{\alpha_i}]_q & \beta \prec \alpha_i
\end{cases}, \qquad
& f_\gamma = \begin{cases}
[f_\beta, f_{\alpha_i}]_q & \alpha_i \prec \beta \\
[f_{\alpha_i}, \, f_\beta]_q & \beta \prec \alpha_i
\end{cases}.
\end{align*}

Now we proceed to the roots $\delta - \gamma$ and $- (\delta - \gamma)$ with $\gamma \in \triangle_+$. We already have the root vectors $e_{\delta - \theta}$ and $f_{\delta - \theta}$ corresponding to the roots $\delta - \theta$ and $-(\delta - \theta)$. Assume that for some number $m$, such that $1 < m \le \hght \theta$, the root vectors $e_{\delta - \gamma}$ and $f_{\delta - \gamma}$ for all $\gamma \in \triangle_+$ with $m \le \hght \gamma \le \hght \theta$ are also defined. Let $\gamma \in \triangle_+$ and $\hght \gamma = m - 1$. It can be shown that for some $i \in I$ the root $\gamma$ can be written as
\begin{equation*}
\gamma = - \alpha_i + \beta,
\end{equation*}
where $\beta \in \triangle_+$ and $\hght \beta = m$. Fixing such a representation we define
\begin{equation*}
e_{\delta - \gamma} = [e_{\alpha_i}, \, e_{\delta - \beta}]_q, \qquad f_{\delta - \gamma} = [f_{\delta - \beta}, \, f_{\alpha_i}]_q.
\end{equation*}

The root vectors corresponding to the roots $\delta$ and $- \delta$ are additionally indexed by the positive roots of $\gothg$\footnote{The same is true for the roots $n \delta$ and $- n \delta$, $k \in \bbN$.} and are defined by the relations
\begin{equation}
e'_{\delta, \, \gamma} = [e_\gamma, \, e_{\delta - \gamma}]_q, \qquad f'_{\delta, \, \gamma} = [f_{\delta - \gamma}, \, f_{\gamma}]_q. \label{epd}
\end{equation}
The remaining definitions are
\begin{gather}
e_{\gamma + n \delta} = [(\gamma | \gamma)]_q^{-1} [e'_{\delta, \, \gamma}, \, e_{\gamma + (n - 1)\delta}]_q, \qquad f_{\gamma + n \delta} = [(\gamma | \gamma)]_q^{-1} [f_{\gamma + (n - 1)\delta}, \, f'_{\delta, \, \gamma}]_q, \label{cweb} \\
e_{(\delta - \gamma) + n \delta} = [(\gamma | \gamma)]_q^{-1} [e_{(\delta - \gamma) + (n - 1)\delta}, \, e'_{\delta, \, \gamma}]_q, \\
f_{(\delta - \gamma) + n \delta} = [(\gamma | \gamma)]_q^{-1} [f'_{\delta, \, \gamma}, \, f_{(\delta - \gamma) + (n - 1)\delta}]_q, \\
e'_{n \delta, \, \gamma} = [e_{\gamma + (n - 1)\delta}, \, e_{\delta - \gamma}]_q, \qquad f'_{n \delta, \, \gamma} = [f_{\delta - \gamma}, \, f_{\gamma + (n - 1)\delta}]_q. \label{cwee}
\end{gather}
In fact, only the root vectors $e_{n \delta, \, \alpha_i}$ and $f_{n \delta, \, \alpha_i}$, $i \in I$, are independent and needed for the construction of the Poincar\'e--Birkhoff--Witt basis.

We will need also another set of root vectors corresponding to the roots $n \delta$ and $- n \delta$, $n \in \bbN$. They are defined by the equations
\begin{gather}
- \kappa_q \, e_{\delta, \gamma}(u) = \log(1 - \kappa_q \, e'_{\delta, \, \gamma}(u)), \label{edg} \\
\kappa_q \, f_{\delta, \gamma}(u^{-1}) = \log(1 + \kappa_q \, f'_{\delta, \, \gamma}(u^{-1})), \label{fdg}
\end{gather}
where we used the generating functions
\begin{align*}
& e'_{\delta, \, \gamma}(u) = \sum_{k = 1}^\infty e'_{n \delta, \, \gamma} \, u^n, && e_{\delta, \, \gamma}(u) = \sum_{n = 1}^\infty e_{n \delta, \, \gamma} \, u^n, \\
& f'_{\delta, \, \gamma}(u^{-1}) = \sum_{n = 1}^\infty f'_{n \delta, \, \gamma} \, u^{- n}, && f_{\delta, \, \gamma}(u^{-1}) = \sum_{n = 1}^\infty f_{n \delta, \, \gamma} \, u^{- n}
\end{align*}
defined as formal power series.

\subsection{Second Drinfeld's realization}

The quantum loop algebra $\uqlg$ has another realization \cite{Dri87, Dri88} as the algebra with generators $\xi^\pm_{i, \, n}$ with $i \in I$ and $n \in \bbZ$, $q^x$ with $x \in \gothh$, and $\chi_{i, \, n}$ with $i \in I$ and $n \in \bbZ \setminus \{0\}$. They satisfy the relations
\begin{gather*}
q^0 = 1, \qquad q^{x_1} q^{x_2} = q^{x_1 + x_2}, \\
[\chi^{\mathstrut}_{i, \, n}, \, \chi^{\mathstrut}_{j, \, m}] = 0, \qquad q^x \chi_{j, \, n} = \chi_{j, \, n} \, q^x,  \\
q^x \xi^\pm_{i, \, n} q^{- x} = q^{\pm \langle \alpha_i, \, x \rangle} \xi^\pm_{j, \, n}, \qquad [\chi^{\mathstrut}_{i, \, n}, \, \xi^\pm_{j, m}] = \pm \frac{1}{n} [n \, b^{\mathstrut}_{i j}]^{\mathstrut}_{q} \, \xi^\pm_{j, \, n + m}, \\
\xi^\pm_{i, \, n + 1} \xi^\pm_{j, \, m} - q^{\pm b_{i j}} \, \xi^\pm_{j, \, m} \, \xi^\pm_{i, \, n + 1} = q^{\pm b_{i j}} \, \xi^\pm_{i, \, n} \, \xi^\pm_{j, \, m + 1} - \xi^\pm_{j, \, m + 1} \xi^\pm_{i, \, n}, \\
[\xi^+_{i, \, n}, \, \xi^-_{j, \, m}] = \delta_{i j} \, \frac{\phi^+_{i, \, n + m} - \phi^-_{i, \, n + m}}{q_i^{\mathstrut} - q_i^{-1}}
\end{gather*}
and the Serre relations whose explicit form is not important for us. The quantities $\phi^\pm_{i, \, n}$, $i \in I$, $n \in \bbZ$, are determined by the formal power series
\begin{equation}
\sum_{n = 0}^\infty \phi^\pm_{i, \, \pm n} u^{\pm n} = q_i^{\pm h_i} \exp \left( \pm \kappa_q \sum_{n = 1}^\infty \chi_{i, \, \pm n} u^{\pm n} \right) \label{phipm}
\end{equation}
and by the conditions
\begin{equation*}
\phi^+_{i, \, n} = 0, \quad n < 0, \qquad \phi^-_{i, \, n} = 0, \quad n > 0.
\end{equation*}

The generators of the second Drinfeld's realization can be related to the Cartan--Weyl generators in the following way \cite{KhoTol93, KhoTol94}. The Drinfeld--Jimbo generators $q^x$ and the generators $q^x$ of the second Drinfeld's realizations are the same, except that in the former case $x \in \tgothh$ and in the latter case $x \in \gothh \subset \tgothh$. For the generators $\xi^\pm_{i, \, n}$ and $\chi_{i, \, n}$ of the second Drinfeld's realization we have
\begin{gather}
\xi^+_{i, \, n} = \begin{cases}
(-1)^n o_i^n e_{\alpha_i + n \delta} & n \ge 0 \\
(-1)^{n + 1} o_i^n q_i^{-h_i} f_{(\delta - \alpha_i) - (n + 1)\delta} & n < 0
\end{cases}, \label{ksipin} \\
\xi^-_{i, \, n} = \begin{cases}
(-1)^n o_i^{n + 1} e_{(\delta - \alpha_i) + (n - 1) \delta} \, q_i^{h_i} & n > 0 \\
(-1)^n o_i^n f_{\alpha_i - n \delta} & n \le 0
\end{cases}, \label{ksimin} \\
\chi_{i, \, n} = \begin{cases}
(-1)^{n + 1} o_i^n e_{n \delta, \, \alpha_i} & n > 0 \\
(-1)^{n + 1} o_i^n f_{- n\delta, \, \alpha_i} & n < 0
\end{cases}, \label{chiin}
\end{gather}
where for each $i \in I$ the number $o_i$ is either $+1$ or $-1$, such that $o_i = - o_j$ whenever $a_{i j} < 0$. It follows from (\ref{edg}), (\ref{fdg}), (\ref{phipm}) and (\ref{chiin}) that
\begin{gather*}
\phi^+_{i, \, n} = \begin{cases}
(-1)^{n + 1} o_i^n \kappa_q \, q_i^{h_i} e'_{n \delta, \, \alpha_i} & n > 0 \\
q_i^{h_i} & n = 0
\end{cases}, \\
\phi^-_{i, \, n} = \begin{cases}
q_i^{-h_i} & n = 0 \\
(-1)^n o_i^n \kappa_q \, q_i^{-h_i} f'_{- n \delta, \, \alpha_i} & n < 0
\end{cases}.
\end{gather*}
Defining the generating functions $\phi^+_i(u)$ and $\phi^-_i(u)$ as
\begin{equation*}
\phi^+_i(u) = \sum_{n = 0}^\infty \phi^+_{i, \, n} u^n, \qquad \phi^-_i(u^{-1}) = \sum_{n = 0}^\infty \phi^-_{i, \, -n} u^{- n},
\end{equation*}
we also obtain
\begin{align}
& \phi^+_i(u) = q_i^{h_i} \big(1 - \kappa_q \, e'_{\delta, \, \alpha_i}(- o_i u)\big), \label{phipiu} \\ 
& \phi^-_i(u^{-1}) = q_i^{-h_i} \big(1 + \kappa_q \, f'_{\delta, \, \alpha_i}(- o_i u^{-1})\big). \label{phimiu}
\end{align}

\section{\texorpdfstring{Highest $\ell$-weight representations of quantum loop algebras}{Highest l-weight representations of quantum loop algebras}}

\subsection{General information}

The terminology used for $\uqlg$-modules \cite{MukYou14} is very similar to the terminology used for $\uqg$-modules \cite{ChaPre91, KliSch97}. We adopt it to the definition of a quantum loop algebra used in the present paper.

A $\uqlg$-module $V$ is said to be a {\em weight module\/} if
\begin{equation}
V = \bigoplus_{\lambda \in \tgothh^*}  V_\lambda, \label{vvl}
\end{equation}
where
\begin{equation*}
V_\lambda = \{v \in V \mid q^x v = q^{\langle \lambda, \, x \rangle} v \mbox{ for any } x \in \tgothh \}.
\end{equation*}
The space $V_\lambda$ is called the {\em weight space\/} of weight $\lambda$, and a nonzero element of $V_\lambda$ is called a {\em weight vector\/} of weight $\lambda$. We say that $\lambda \in \gothh^*$ is a {\em weight\/} of $V$ if $V_\lambda \ne \{0\}$. It follows from (\ref{qnuc}) that any weight $\lambda$ of a $\uqlg$-module satisfies the relation (\ref{lambdac}), hence, it can be identified with a unique element of $\gothh^*$.

We say that a $\uqlg$-module $V$ is in the {\em category\/} $\calO$ if
\begin{itemize}
\item[(i)] V is a weight module all of whose weight spaces are finite dimensional;
\item[(ii)] there exists a finite number of elements $\mu_1, \ldots, \mu_s \in \gothh^*$ such that every weight of $V$ belongs to the set
\begin{equation*}
\bigcup_{i = 1}^s \, \{\mu \in \gothh^* \mid \mu \leq \mu_i \},
\end{equation*}
where $\leq$ is the usual partial order in $\gothh^*$.
\end{itemize}
 A $\uqlg$-module $V$ in the category $\calO$ is called a {\em highest weight module\/} with {\em highest weight\/} $\lambda$ if there exists a weight vector $v \in V$ of weight $\lambda$ such that
\begin{equation*}
e_i v = 0
\end{equation*}
for all $i \in I$, and
\begin{equation*}
V = \uqlg v.
\end{equation*}
The vector with the above properties is unique up to a scalar factor. We call it the {\em highest weight vector\/} of $V$.

By definition, for any $\uqlg$-module $V$ in the category $\calO$ we have the decomposition (\ref{vvl}). We can refine it in the following way. Define an {\em $\ell$-weight\/} as a set
\begin{equation*}
\bm \Psi = \{ \Psi^+_{i, \, n} \in \bbC \mid i \in I, \, n \in \bbZ_+ \} \cup \{ \Psi^-_{i, \,- n} \in \bbC \mid i \in I, \, n \in \bbZ_+ \}
\end{equation*}
such that
\begin{equation}
\Psi^+_{i, \, 0} \, \Psi^-_{i, \, 0} = 1. \label{pppmo*}
\end{equation}
Now we have
\begin{equation}
V_\lambda = \bigoplus_{\bm \Psi} V_{\bm \Psi}, \label{vvp}
\end{equation}
where $V_{\bm \Psi}$ is a subspace of $V_\lambda$ such that for any $v$ in $V_{\bm \Psi}$ there is $p \in \bbN$ such that
\begin{equation}
(\phi^\pm_{i, \, \pm n} - \Psi^\pm_{i, \, \pm n})^p v = 0 \label{pppmo}
\end{equation}
for all $i \in I$ and $n \in \bbZ_+$. The space $V_{\bm \Psi}$ is called the {\em $\ell$-weight space\/} of $\ell$-weight $\bm \Psi$. We say that $\bm \Psi$ is an {\em $\ell$-weight\/} of $V$ if $V_{\bm \Psi} \ne \{0\}$. A nonzero element $v \in V_{\bm \Psi}$ such that
\begin{equation*}
\phi^\pm_{i, \, \pm n} v = \Psi^\pm_{i, \, \pm n} v
\end{equation*}
for all $i \in I$ and $n \in \bbZ_+$ is said to be an {\em $\ell$-weight vector\/} of $\ell$-weight $\bm \Psi$. Every nontrivial $\ell$-weight space contains an $\ell$-weight vector. It is clear that $V_{\bm \Psi}$ in the decomposition (\ref{vvp}) is nontrivial only if
\begin{equation*}
\Psi^\pm_{i, \, 0} = q_i^{\pm \langle \lambda, \, h_i \rangle}.
\end{equation*} 

A $\uqlg$-module $V$ in the category $\calO$ is called a {\em highest $\ell$-weight module\/} with {\em highest $\ell$-weight\/} $\bm \Psi$ if there exists an $\ell$-weight vector $v \in V$ of $\ell$-weight $\bm \Psi$ such that
\begin{equation*}
\xi^+_{i, \, n} v = 0
\end{equation*}
for all $i \in I$ and $n \in \bbZ$, and
\begin{equation*}
V = \uqlg v.
\end{equation*}
The vector with the above properties is unique up to a scalar factor. We call it the {\em highest $\ell$-weight vector\/} of $V$.

Given $\ell$-weight $\bm \Psi$, define two sets of generating functions 
$\Psi^+_i(u)$ and $\Psi^-_i(u^{-1})$ as 
\begin{equation*}
\Psi^+_i(u) = \sum_{n \in \bbZ_+} \Psi^+_{i, \, n} u^n, \qquad \Psi^-_i(u^{-1}) = \sum_{n \in \bbZ_+} \Psi^-_{i, \, - n} u^{- n}.
\end{equation*}
When it is convenient, we will identify $\bm \Psi$ with the set $\{\Psi^+_i(u), \, \Psi^-_i(u^{-1})\}_{i \in I}$. An $\ell$-weight $\bm \Psi$ is called {\em rational\/} if for some non-negative integers $p_i$, $i \in I$, and complex numbers $a_{i k}$, $b_{i k}$, $i \in I$, $0 \le k \le p_i$, one has
\begin{align}
& \Psi^+_i(u) = \frac{a_{i p_i} u^{p_i} + a_{i, \, p_i - 1} u^{p_i - 1} + \cdots + a_{i 0}}{b_{i p_i} u^{p_i} + b_{i, \, p_i - 1} u^{p_i - 1} + \cdots + b_{i 0}}, \label{psipi} \\
& \Psi^-_i(u^{-1}) = \frac{a_{i p_i} + a_{i, \, p_i - 1} u^{- 1} + \cdots + a_{i 0} u^{- p_i}}{b_{i p_i} + b_{i, \, p_i - 1} u^{- 1} + \cdots + b_{i 0} u^{- p_i}}. \label{psimi}
\end{align}
Here $a_{i p_i}$, $a_{i 0}$, $b_{i p_i}$, $b_{i 0}$ must be nonzero and such that
\begin{equation*}
\frac{a_{i p_i}}{b_{i p_i}} \frac{a_{i 0}}{b_{i 0}} = 1.
\end{equation*}
This relation guaranties the validity of equation (\ref{pppmo*}).

The remarkable fact is that for any rational $\ell$-weight $\bm \Psi$ there is an irreducible highest $\ell$-weight $\uqlg$-module $L(\bm \Psi)$ with highest $\ell$-weight $\bm \Psi$  which is unique up to an isomorphism, and any irreducible $\uqlg$-module in the category $\calO$ is a highest $\ell$-weight module with a rational highest $\ell$-weight. In other words, there is a bijection between the rational $\ell$-weights and the equivalence classes of the irreducible $\uqlg$-modules in the category $\calO$. Furthermore, all $\ell$-weights of a $\uqlg$-module in the category $\calO$ are rational.

For any rational $\ell$-weights $\bm \Psi$ and $\bm \Psi'$ define the rational $\ell$-weight $\bm \Psi \bm \Psi'$ as the set $\{\Psi^+_i(u) \Psi'^+_i(u), \, \Psi^-_i(u^{-1}) \Psi'^-_i(u^{-1})\}_{i \in I}$. One can show that the submodule of the tensor product $L(\bm \Psi) \otimes L(\bm \Psi')$ generated by the tensor product of the highest $\ell$-weight vectors is a highest $\ell$-weight module with highest $\ell$-weight $\bm \Psi \bm \Psi'$. In particular, $L(\bm \Psi \bm \Psi')$ is a subquotient of $L(\bm \Psi) \otimes L(\bm \Psi')$.

\subsection{Jimbo's homomorphism}

In the present paper we deal with quantum loop algebras associated with Kac--Moody algebras $\widehat \gothg$ defined by the generalized Cartan matrices $\widehat A = A^{(1)}_l$. The corresponding generalized Cartan matrices of finite type are $A_l$ and the corresponding finite dimensional Kac--Moody algebras are isomorphic to the Lie algebras $\gothsl_{l + 1}$. Thus, we deal with the quantum loop algebras $\mathrm U_q(\mathcal L(\mathfrak{sl}_{l + 1}))$. The usual way to construct highest $\ell$-weight representations in the case under consideration is to use the Jimbo's homomorphism. It can be defined as a homomorphism from $\mathrm U_q(\mathcal L(\mathfrak{sl}_{l + 1}))$ to the quantum group $\mathrm U_q(\mathfrak{sl}_{l + 1})$, however, it is convenient to define it as a homomorphism to the quantum group $\mathrm U_q(\mathfrak{gl}_{l + 1})$. Let us recall the definition of the quantum groups $\mathrm U_q(\mathfrak{sl}_{l + 1})$ and $\mathrm U_q(\mathfrak{gl}_{l + 1})$.

It is common to denote the generators of quantum groups associated with finite dimensional Kac--Moody algebras by upper case letters. Following this custom, we say that the quantum group $\mathrm U_q(\mathfrak{sl}_{l + 1})$ is defined by the generators $E_i$, $F_i$, $i \in I$, and $q^X$, where $X$ belongs to the Cartan subalgebra $\gothh$ of $\mathfrak{sl}_{l + 1}$.\footnote{For consistency we denote the Cartan generators of a finite dimensional Lie algebra by~$H_i$.} These generators satisfy the defining relations
\begin{gather}
q^0 = 1, \qquad q^{X_1} q^{X_2} = q^{X_1 + X_2}, \label{djrfa} \\
q^X E_i \, q^{-X} = q^{\langle \alpha_i, \, X \rangle} E_i, \qquad q^X F_i \, q^{-X} = q^{- \langle \alpha_i, \, X \rangle} F_i, \\
[E_i, \, F_j] = \delta_{i j} \, \frac{q_i^{H_i} - q_i^{-H_i}}{q_i^{\mathstrut} - q_i^{-1}}, \\
\sum_{n = 0}^{1 - a_{i j}} (-1)^n E_i^{(1 - a_{i j} - n)} E^{\mathstrut}_j \, E_i^{(n)} = 0, \qquad \sum_{n = 0}^{1 - a_{i j}} (-1)^n F_i^{(1 - a_{i j} - n)} F^{\mathstrut}_j \, F_i^{(n)} = 0, \label{djrfd}
\end{gather}
where $E_i^{(n)} = E_i^n / [n]_{q_i}!$, $F_i^{(n)} = F_i^n / [n]_{q_i}!$. Now $\alpha_i$ are the simple roots of $\gothsl_{l + 1}$ defined as
\begin{equation}
\langle \alpha_i, \, H_j \rangle = a_{j i}, \label{aihj}
\end{equation}
where $a_{i j}$ are the entries of the Cartan matrix of $\gothsl_{l + 1}$. 
In the case under consideration, the generalized Cartan matrix is symmetric. 
Therefore, all integers $d_i$ are just $1$, and so, $q_i = q$ everywhere in (\ref{djrfa})--(\ref{djrfd}).

The Lie algebra $\gothgl_{l + 1}$ can be considered as a trivial central extension of the Lie algebra $\gothsl_{l + 1}$ by a one-dimensional centre $\bbC \, K$, so that as the Cartan subalgebra $\gothk$ of $\gothgl_{l + 1}$ one can take the space
\begin{equation*}
\gothk = \gothh \oplus \bbC \, K = \Big( \bigoplus_{i \in I} H_i \Big) \oplus \bbC \, K.
\end{equation*}
It is convenient together with the basis of the Cartan subalgebra of $\gothgl_{l + 1}$ formed by the elements $H_i$, $i \in I$, and $K$ to use the basis formed by the elements $K_i$, $i = 1, \ldots, l + 1$, defined so that
\begin{equation*}
H_i = K_i - K_{i + 1}, \quad i = 1, \ldots, l, \qquad K = \sum_{i = 1}^{l + 1} K_i.
\end{equation*}
It is not difficult to demonstrate that
\begin{equation*}
K_i = \frac{1}{l + 1} \Bigl(K - \sum_{j = 1}^{i - 1} j \, H_j + \sum_{j = i}^l (l + 1 - j) H_j \Bigr).
\end{equation*}
We define the quantum group $\mathrm U_q(\mathfrak{gl}_{l + 1})$ as a trivial central extension of $\mathrm U_q(\mathfrak{sl}_{l + 1})$ performed by adding the generators $q^{\nu K}$, $\nu \in \bbC^\times$. The defining relations of $\mathrm U_q(\mathfrak{gl}_{l + 1})$ have the form (\ref{djrfa})--(\ref{djrfd}), where $X, X_1, X_2 \in \gothk$, and the simple roots $\alpha_i \in \gothk^*$, $i \in I$, are defined by the relations (\ref{aihj}) supplemented by the equation
\begin{equation*}
\langle \alpha_i, \, K \rangle = 0.
\end{equation*}

Let $\lambda$ be an element of $\gothk^*$. We identify $\lambda$ with the set of its components $(\lambda_1, \ldots, \lambda_{l + 1})$ with respect to the dual basis of the basis $\{K_i\}$. In fact we have
\begin{equation*}
\lambda_i = \lambda(K_i).
\end{equation*}
We denote by $\widetilde V^\lambda$ the infinite dimensional highest weight $\mathrm U_q(\mathfrak{gl}_{l + 1})$-module with the highest weight vector $v_0$. By definition, we have
\begin{equation}
q^X v_0 = q^{\langle \lambda, \, X \rangle} v_0, \qquad E_i v_0 = 0, \quad i \in I. \label{vmr}
\end{equation}
Note that the first equation of (\ref{vmr}) is equivalent to
\begin{equation*}
q^{\nu K_i} v_0 = q^{\nu \lambda_i} v_0, \qquad i = 1, \ldots, l + 1, \quad \nu \in \bbC,
\end{equation*}
where
\begin{equation*}
\lambda_i = \langle \lambda, \, K_i \rangle.
\end{equation*}
Below we identify $\lambda$ with the ordered set of the numbers $\lambda_i$. It is known that when $\lambda_i - \lambda_{i + 1}$ for all $i \in I$ are non-negative integers the module $\widetilde V^\lambda$ is reducible. Here the quotient of $\widetilde V^\lambda$ by the unique maximal submodule is finite dimensional. We denote this finite dimensional $\mathrm U_q(\mathfrak{gl}_{l + 1})$-module by $V^\lambda$. The representations of $\uqgllpo$ corresponding to the modules $\widetilde V^\lambda$ and $V^\lambda$ are denoted by $\widetilde \pi^\lambda$ and $\pi^\lambda$.

As we noted above, to construct representations of $\uqlsllpo$ we are going to use the Jimbo's homomorphism $\varepsilon \colon \uqlsllpo \to \uqgllpo$ introduced in the paper \cite{Jim86a}. We will give the explicit form of $\varepsilon$ for $l = 1, 2$ below. If $\pi$ is a representation of $\uqgllpo$, then $\pi \circ \varepsilon$ is a representation of $\uqlsllpo$. In particular, starting with the representations $\widetilde \pi^\lambda$ and $\pi^\lambda$ described above, we define the representations
\begin{equation}
\widetilde \varphi^\lambda = \widetilde \pi^\lambda \circ \varepsilon, \qquad \varphi^\lambda = \pi^\lambda \circ \varepsilon. \label{wpp}
\end{equation}
Slightly abusing notation, we denote the corresponding $\uqlsllpo$-modules by $\widetilde V^\lambda$ and $V^\lambda$. The $\uqlsllpo$-modules $\widetilde V^\lambda$ and $V^\lambda$ are highest $\ell$-weight modules in the category $\calO$. Here the highest weight vectors of $\widetilde V^\lambda$ and $V^\lambda$ considered as $\uqgllpo$-modules are the highest $\ell$-weight vectors of $\widetilde V^\lambda$ and $V^\lambda$ considered as $\uqlsllpo$-modules.

There is an evident automorphism $\sigma$ of $\uqlsllpo$ defined by the equation
\begin{align}
& \sigma(e_0) = e_1, && \sigma(e_1) = e_2, && \ldots && \sigma(e_l) = e_0, \label{sigmab} \\
& \sigma(f_0) = f_1, && \sigma(f_1) = f_2, && \ldots && \sigma(f_l) = f_0, \\
& \sigma(q^{h_0}) = q^{h_1}, && \sigma(q^{h_1}) = q^{h_2}, && \ldots && \sigma(q^{h_l}) = q^{h_0}. \label{sigmae}
\end{align}
It is evident that $\sigma^{l + 1}$ is the identity transformation. One can consider the representations of $\uqlsllpo$ obtained from the representations $\widetilde \varphi^\lambda$ and $\varphi^\lambda$ defined by (\ref{wpp}) via twisting by powers of $\sigma$. It is clear that in this way we obtain representations that are not highest $\ell$-weight representations. Moreover, from the point of view of the theory of quantum integrable systems these representations are not very interesting because they give practically the same transfer matrices as the initial representations $\widetilde \varphi^\lambda$ and $\varphi^\lambda$. However, considering representations of the Borel subalgebras of $\uqlsllpo$ we use the automorphism $\sigma$ to obtain new interesting representations, see section \ref{s:rbs}.

Another evident automorphism $\tau$ of $\uqlsllpo$ is defined as
\begin{gather}
\tau(e_0) = e_0, \qquad \tau(e_i) = e_{l - i + 1}, \qquad \tau(f_0) = f_0, \qquad \tau(f_i) = f_{l - i + 1}, \qquad i \in I, \label{taub} \\
\tau(q^{h_0}) = q^{h_0}, \qquad \tau(q^{h_i}) = q^{h_{l - i + 1}}, \qquad i \in I. \label{taue}
\end{gather}
It is clear $\tau^2$ is the identity transformation. The twisting of the representations $\widetilde \varphi^\lambda$ and $\varphi^\lambda$ by the automorphism $\tau$ leads to new useful representations of $\uqlsllpo$ and its Borel subalgebras. However, this automorphism is trivial in the case of $\uqlslii$.

\subsection{\texorpdfstring{Case of $\gothg = \slii$}{Case of g = sl2}}

In this case the Jimbo's homomorphism is defined by the relations\footnote{Below in the case of $\gothg = \slii$ instead of $E_1$, $F_1$ and $H_1$ we write just $E$, $F$ and $H$. Similarly, we write $e'_{n \delta}$ and $f'_{n \delta}$ instead of $e'_{n \delta, \, \alpha_1}$ and $f'_{n \delta, \, \alpha_1}$ etc. We also assume that $o_1 = 1$.}
\begin{gather*}
\varepsilon(e_0) = F \, q^{K_1 + K_2}, \qquad \varepsilon(e_1) = E, \qquad \varepsilon(f_0) = E \, q^{- K_1 - K_2}, \qquad \varepsilon(f_1) = F, \\
\varepsilon(q^{\nu h_0}) = q^{\nu(K_2 - K_1)}, \qquad  \varepsilon(q^{\nu h_1}) = q^{\nu (K_1 - K_2)}.
\end{gather*}
Via this automorphism any $\uqglii$-module can be considered as a $\uqlslii$-module. In this section we deal with the highest weight $\uqglii$-modules $\widetilde V^\lambda$ defined in the previous section. Given $\lambda = (\lambda_1, \, \lambda_2) \in \gothk^*$, the vectors
\begin{equation*}
v_m = F^m v_0,
\end{equation*}
where $m \in \bbZ_+$, form a basis of $\widetilde V^\lambda$. The action of the generators of $\uqglii$ on the elements of this basis is described by the formulae
\begin{gather}
q^{\nu K_1} v_m = q^{\nu(\lambda_1 - m)} v_m, \qquad q^{\nu K_2} v_m = q^{\nu(\lambda_2 + m)} v_m, \label{vsliia} \\
E v_m = [m]_q [\lambda_1 - \lambda_2 - m + 1]_q v_{m - 1}, \qquad F v_m = v_{m + 1}. \label{vsliib}
\end{gather}

To define Cartan--Weyl generators we use the following normal order of $\widehat \triangle$:
\begin{equation*}
\alpha, \, \alpha + \delta, \, \ldots, \, \alpha + k \delta, \, \ldots, \, \delta, \, 2 \delta, \,
\ldots, \, k \delta, \, \ldots, \, \ldots, \, (\delta - \alpha) + k \delta, \, \ldots, \,
(\delta - \alpha) + \delta, \, \delta - \alpha,
\end{equation*}
see the paper \cite{TolKho92}. Defining the generating function
\begin{equation*}
\bbE'_\delta(u) = \sum_{n = 1}^\infty \varepsilon(e'_{n \delta}) u^n,
\end{equation*}
we obtain from (\ref{phipiu}) that
\begin{equation*}
\varepsilon(\phi^+(u)) = q^{K_1 - K_2} \big(1 - \kappa_q \, \bbE'_\delta(- u)\big).
\end{equation*}
Using formulas from \cite{NirRaz14} or applying the method of the paper \cite{Raz13} to the case of $\uqlslii$, we see that if we denote
\begin{align*}
& \bbN'_{11}(u) = 1 - u q^{2 K_1}, \qquad \hspace{2.1em} \bbN'_{12} = - \kappa_q q^{-1} F \, q^{K_1 + K_2}, \\
& \bbN'_{21} = - \kappa_q E, \qquad \hspace{5.2em} \bbN'_{22}(u) = 1 - u q^{2 K_2}, \\
& \hspace{3.8em} \bbN''_{22}(u) = \bbN'_{22}(u) - u \bbN'_{21} \bbN'^{-1}_{11}(u) \bbN'_{12},
\end{align*}
then we obtain
\begin{equation*}
1 - \kappa_q \, \bbE'_{\delta}(u) = \bbN'^{-1}_{11}(-q^2 u) \bbN''_{22}(-q^2 u).
\end{equation*}
Hence, as follows from (\ref{phipiu}), we have
\begin{equation*}
\varepsilon(\phi^+(u)) = q^{K_1 - K_2} \, \bbN'^{-1}_{11}( q^2 u) \, \bbN''_{22}(q^2 u),
\end{equation*}
and using (\ref{vsliia}) and (\ref{vsliib}), we come to
\begin{equation}
\phi^+(u) v_m = q^{\lambda_1 - \lambda_2 - 2 m} \, \frac{(1 - q^{2 \lambda_1 + 2} \, u)(1 - q^{2 \lambda_2} \, u)}{(1 - q^{2 \lambda_1 + 2 - 2 m} \, u)(1 - q^{2 \lambda_1 - 2 m} \, u)} \, v_m. \label{epsphipv}
\end{equation}

Introducing the generating function
\begin{equation*}
\bbF'_\delta(u^{- 1}) = \sum_{n = 1}^\infty \varepsilon(f'_{n \delta}) u^{- n},
\end{equation*}
we obtain from (\ref{phimiu}) the equation
\begin{equation*}
\varepsilon(\phi^-(u^{- 1})) = q^{K_2 - K_1} \big(1 + \kappa_q \, \bbF'_\delta(- u^{- 1})\big).
\end{equation*}
Using again formulas from \cite{NirRaz14} or applying the method of the paper \cite{Raz13} to the case of $\uqlslii$, we see that if one defines
\begin{align*}
& \bbO'_{11}(u^{- 1}) = 1 - u^{-1} q^{- 2 K_1}, \qquad \hspace{2.em} \bbO'_{12} = \kappa_q F, \hspace{1.9em} \\
& \bbO'_{21} = \kappa_q q E \, q^{- K_1 - K_2}, \qquad \hspace{4.6em} \bbO'_{22}(u^{- 1}) = 1 - u^{-1} q^{-2 K_2}, \\
& \hspace{3.em} \bbO''_{22}(u^{- 1}) = \bbO'_{22}(u^{- 1}) - u^{-1} \bbO'_{21} \bbO'^{-1}_{11}(u^{- 1}) \bbO'_{12},
\end{align*}
then
\begin{equation*}
1 + \kappa_q \, \bbF'_{\delta}(u^{- 1}) = \bbO''_{22}(-q^{- 2} u^{- 1}) \bbO'^{-1}_{11}(-q^{- 2} u^{- 1}).
\end{equation*}
Hence, as follows from (\ref{phipiu}), we have
\begin{equation*}
\varepsilon(\phi^-(u^{-1})) = q^{- (K_1 - K_2)} \bbO''_{22}(q^{- 2} u^{- 1}) \bbO'^{-1}_{11}(q^{- 2} u^{- 1})
\end{equation*}
Using again (\ref{vsliia}) and (\ref{vsliib}), we see that
\begin{equation}
\phi^-(u^{-1}) v_m = q^{- (\lambda_1 - \lambda_2) + 2 m} \frac{(1 - q^{- 2 \lambda_1 - 2} u^{- 1})(1 - q^{- 2 \lambda_2} u^{- 1})}{(1 - q^{- 2 \lambda_1 - 2 + 2 m} u^{- 1})(1 - q^{- 2 \lambda_1 + 2 m} u^{- 1})} \, v_m. \label{epsphimv}
\end{equation}
in agreement with (\ref{psipi}) and (\ref{psimi}).

\subsection{\texorpdfstring{Case of $\gothg = \sliii$}{Case of g = sl3}} \label{s:fsl3}

In the case under consideration, to define the Jimbo's homomorphism we denote
\begin{equation*}
E_3 = E_1 E_2 - q \, E_2 E_1, \qquad F_3 = F_2 F_1 - q^{-1} F_1 F_2.
\end{equation*}
Now, the Jimbo's homomorphism $\varepsilon$ is determined by the relations
\begin{align}
& \varepsilon(q^{\nu h_0}) = q^{\nu(K_3 - K_1)}, && \varepsilon(q^{\nu h_1}) = q^{\nu (K_1 - K_2)}, && \varepsilon(q^{\nu h_2}) = q^{\nu (K_2 - K_3)}, \label{j2h} \\
& \varepsilon(e_0) = F_3 \, q^{K_1 + K_3}, && \varepsilon(e_1) = E_1, && \varepsilon(e_2) = E_2, \label{j2e} \\
& \varepsilon(f_0) = E_3 \, q^{- K_1 - K_3} , && \varepsilon(f_1) = F_1, && \varepsilon(f_2) = F_2. \label{j2f}
\end{align}

Describe the structure of the highest weight $\uqgliii$-modules. Let $\lambda = (\lambda_1, \, \lambda_2, \, \lambda_3)$ be an arbitrary element of the dual space of the standard Cartan subalgebra $\gothk$ of $\gliii$. The highest weight vector $v_0$ of the module $\widetilde V^\lambda$ satisfies the relations
\begin{equation*}
q^{\nu K_i} v_0 = q^{\nu \lambda_i} v_0, \quad i = 1, 2, 3, \qquad E_i v_0 = 0, \quad i = 1, 2.
\end{equation*}
The vectors
\begin{equation}
v_{\bm m} = F_1^{m_1} F_3^{m_2} F_2^{m_3} v_0, \label{vn}
\end{equation}
where $m_1, m_2, m_3 \in \bbZ_+$ and $\bm m = (m_1, \, m_2, \, m_3)$, form a basis of the $\widetilde V^\lambda$. Here it is natural to assume that $v_0$ means $v_{\bm 0} = v_{(0, \, 0, \, 0)}$. One can find that the action of the generators of $\uqgliii$ and the elements $E_3$ and $F_3$ on the basis vectors $v_{\bm m}$ is described by the formulas\footnote{For any $k_1, k_2, k_3 \in \bbZ$ we use the notation $\bm m + k_1 \bm \epsilon_1 + k_2 \bm \epsilon_2 + k_3 \bm \epsilon_3 = (m_1 + k_1, \, m_2 + k_2, \, m_3 + k_3)$.}
\begin{align*}
& q^{\nu K_1} v_{\bm m} = q^{\nu(\lambda_1 - m_1 - m_2)} v_{\bm m}, \quad q^{\nu K_2} = q^{\nu(\lambda_2 + m_1 - m_3)} v_{\bm m}, \quad q^{\nu K_3} v_{\bm m} = q^{\nu(\lambda_3 + m_2 + m_3)} v_{\bm m}, \\
& F_1 v_{\bm m} = v_{\bm m + \bm \epsilon_1}, \quad F_2 v_{\bm m} =  q^{- m_1 + m_2} v_{\bm m + \bm \epsilon_3} + [m_1]_q v_{\bm m - \bm \epsilon_1 + \bm \epsilon_2}, \quad F_3 v_{\bm m} = q^{m_1} v_{\bm m + \bm \epsilon_2}, \\
& E_1 v_{\bm m} = [\lambda_1 - \lambda_2 - m_1 - m_2 + m_3 + 1]_q [m_1]_q v_{\bm m - \bm \epsilon_1} - q^{- \lambda_1 + \lambda_2 + m_2 - m_3 - 2} [m_2]_q v_{\bm m - \bm \epsilon_2 + \bm \epsilon_3}, \\
& E_2 v_{\bm m} = [\lambda_2 - \lambda_3 - m_3 + 1]_q \, [m_3]_q v_{\bm m - \bm \epsilon_3} + q^{\lambda_2 - \lambda_3 - 2 m_3} [m_2]_q v_{\bm m + \bm \epsilon_1 - \bm \epsilon_2}, \\
& E_3 v_{\bm m} = q^{- m_1} [\lambda_1 - \lambda_3 - m_1 - m_2 - m_3 + 1]_q [m_2]_q v_{\bm m - \bm \epsilon_2} \notag \\
& \hspace{8.em} {} - q^{\lambda_1 - \lambda_2 - m_1 - m_2 + m_3 + 1} [\lambda_2 - \lambda_3 - m_3 + 1]_q [m_1]_q [m_3]_q v_{\bm m - \bm \epsilon_1 - \bm \epsilon_3},
\end{align*}
see, for example, the paper \cite{BooGoeKluNirRaz14b}. The basis of a finite dimensional module $V^\lambda$ is a certain subset of the basis formed by the vectors $v_{\bm m}$. Here the action of the generators is defined again by the above relations supplied with the condition of vanishing of the vectors which are outside of the basis.

In the present case, to define Cartan--Weyl generators we use the following normal order of $\widehat \triangle$:
\begin{gather*}
\alpha_1, \, \alpha_1 + \alpha_2, \, \alpha_1 + \delta, \, \alpha_1 + \alpha_2 + \delta, \, \ldots, \, \alpha_1 +
k \delta, \, \alpha_1 + \alpha_2 + k \delta, \, \ldots, \\* \alpha_2, \, \alpha_2 + \delta, \, \ldots, \,
\alpha_2 + k \delta, \, \ldots, \delta, \, 2 \delta, \, \ldots, \, k \delta, \, \ldots, \ldots, \,
(\delta - \alpha_2) + k \delta, \, \ldots, \, \delta - \alpha_2, \\* \ldots, (\delta - \alpha_1) +
k \delta, \, (\delta - \alpha_1 - \alpha_2) + k \delta, \, \ldots, \, \delta - \alpha_1, \,
\delta - \alpha_1 - \alpha_2,
\end{gather*}
see the paper \cite{BraGouZha95}.

Let us find the $\ell$-weight vectors of the module $\widetilde V^\lambda$. Introduce the generating functions
\begin{equation*}
\bbE'_{\delta, \, \alpha_i}(u) = \sum_{n = 1}^\infty \varepsilon(e'_{n \delta, \, \alpha_i}) u^n.
\end{equation*}
Choosing for definiteness $o_1 = 1$ and $o_2 = -1$, we obtain from (\ref{phipiu}) the equations
\begin{align}
& \varepsilon(\phi^+_1(u)) = q^{K_1 - K_2} \big(1 - \kappa_q \, \bbE'_{\delta, \alpha_1}(- u)\big), \label{eppo} \\
& \varepsilon(\phi^+_2(u)) = q^{K_2 - K_3} \big(1 - \kappa_q \, \bbE'_{\delta, \alpha_2}(u)\big). \label{eppt}
\end{align}
It follows from the results of the paper \cite{Raz13} that if we define
\begin{align}
& \bbN'_{11}(u) = 1 - u \, q^{2 K_1}, && \bbN'_{12} = - \kappa_q \, q^{-1} \, F_1 \, q^{K_1 + K_2}, && \bbN'_{13} = - \kappa_q \, q^{-1} \, F_3 \, q^{K_1 + K_3}, \\
& \bbN'_{21} = - \kappa_q \, E_1 && \bbN'_{22}(u) = 1 - u \, q^{2 K_2}, && \bbN'_{23} = - \kappa_q \, q^{-1} \, F_2 \, q^{K_2 + K_3}, \label{robbmb} \\*
& \bbN'_{31} = - \kappa_q \, E_3, && \bbN'_{32} = - \kappa_q \, E_2  && \bbN'_{33}(u) = 1 - u \, q^{2 K_3}. \label{robbme}
\end{align}
and
\begin{align}
& \bbN''_{22}(u) = \bbN'_{22}(u) - u \, \bbN'_{21} \bbN'^{-1}_{11}(u) \bbN'_{12}, \hspace{1.5em} \bbN''_{23}(u) = \bbN'_{23} - \bbN'_{21} \bbN'^{-1}_{11}(u) \bbN'_{13}, \label{mpa} \\
& \bbN''_{32}(u) = \bbN'_{32} - u \, \bbN'_{31} \bbN_{11}^{\prime -1}(u) \bbN'_{12}, \hspace{3.em} \bbN''_{33}(u) = \bbN'_{33}(u) - u \, \bbN'_{31} \bbN'^{-1}_{11}(u) \bbN'_{13}, \\
& \hspace{8.em} \bbN'''_{33}(u) = \bbN''_{33}(u) - u \, \bbN''_{32}(u) \bbN''^{-1}_{22}(u) \bbN''_{23}(u), \label{mpb}
\end{align}
we obtain 
\begin{align*}
& 1 - \kappa_q \, \bbE'_{\delta, \, \alpha_1}(u) = \bbN'^{-1}_{11}(-q^2 u) \bbN''_{22}(-q^2 u), \\
& 1 - \kappa_q \, \bbE'_{\delta, \, \alpha_2}(u) = \bbN''^{-1}_{22}(q^3 u) \bbN'''_{33}(q^3 u).
\end{align*}
Comparing these equations with (\ref{eppo}) and (\ref{eppt}), we come to the relations
\begin{align}
& \varepsilon(\phi^+_1(u)) = q^{K_1 - K_2} \bbN'^{-1}_{11}(q^2 u) \bbN''_{22}(q^2 u), \label{ezppo} \\
& \varepsilon(\phi^+_2(u)) = q^{K_2 - K_3}\bbN''^{-1}_{22}(q^3 u) \bbN'''_{33}(q^3 u). \label{ezppt}
\end{align} 

It is not difficult to determine that
\begin{multline}
\bbN''_{2 2}(u) v_{\bm m} =  \frac{1}{(1 - q^{2 \lambda_1 - 2 m_1 - 2 m_2 - 2} u)} \\*
\times \Big[ (1 - q^{2 \lambda_1 - 2 m_2} u)  (1 - q^{2 \lambda_2 - 2 m_3 - 2} u) v_m \\* + \kappa_q^2 \, q^{2 \lambda_2 - 2 m_3 - 3} u [m_2]_q \, v_{\bm m + \bm \epsilon_1 - \bm \epsilon_2 + \bm \epsilon_3} \Big], \label{npptt}
\end{multline}
and, therefore, as follows from (\ref{ezppo}) we have
\begin{multline*}
\phi^+_1(u) v_{\bm m} =  \frac{q^{\lambda_1 - \lambda_2 - 2 m_1 - m_2 + m_3}}{(1 - q^{2 \lambda_1 - 2 m_1 - 2 m_2 + 2} u) (1 - q^{2 \lambda_1 - 2 m_1 - 2 m_2} u)} \\
\times \Big[  (1 - q^{2 \lambda_1 - 2 m_2 + 2} u)  (1 - q^{2 \lambda_2 - 2 m_3} u) v_{\bm m} \\ + \kappa_q^2 \, q^{2 \lambda_2 - 2 m_3 - 1} u [m_2]_q \, v_{\bm m + \bm \epsilon_1 - \bm \epsilon_2 + \bm \epsilon_3} \Big].
\end{multline*}
This relation suggests us to look for $\ell$-weight vectors in the form
\begin{equation}
w_{\bm m} = \sum_{k = 0}^{m_2} C_{k, \, \bm m} \, v_{\bm m + k \bm \epsilon_1 - k \bm \epsilon_2 + k \bm \epsilon_3}. \label{wm}
\end{equation}
After some calculations, we see that if we put
\begin{equation}
C_{k, \, \bm m} = (-1)^k \kappa_q^k \, q^{- (k - 1)k/2} \qbinom{m_2}{k}_q \bigg[ \prod_{i = 1}^k (1 - q^{2 \lambda_1 - 2 \lambda_2 - 2 m_2 + 2 m_3 + 2 i + 2}) \bigg]^{-1} \label{ckm}
\end{equation}
for $k = 1,\ldots,m_2$, and $C_{0, \, \bm m} = 1$, then we obtain
\begin{equation*}
\phi^+_1(u) w_{\bm m} = \Psi^+_{1, \, \bm m}(u) w_{\bm m},
\end{equation*}
where
\begin{equation}
\Psi^+_{1, \bm m}(u) = q^{\lambda_1 - \lambda_2 - 2 m_1 - m_2 + m_3} \, \frac{1 - q^{2 \lambda_1 - 2 m_2 + 2} u}{1 - q^{2 \lambda_1 - 2 m_1 - 2 m_2 + 2} u} \, \frac{1 - q^{2 \lambda_2 - 2 m_3} u}{1 - q^{2 \lambda_1 - 2 m_1 - 2 m_2} u}. \label{psip1m}
\end{equation}
It is clear that the vectors $w_{\bm m}$ form a basis of the module $\widetilde V^\lambda$ and the corresponding subset of this basis is a basis of the module $V^\lambda$.

The calculations for the case of $\phi^+_2(u)$ are more complicated. We give only a few intermediate formulas. First note that as follows from (\ref{npptt}) for the action of $\bbN''^{-1}_{22}(u)$ on the basis vectors $v_{\bm m}$ we have the representation
\begin{equation*}
\bbN''^{-1}_{22}(u) v_{\bm m} = \sum_{k = 0}^{m_2} A_{k, \, \bm m}(u) \, v_{\bm m + k \bm \epsilon_1 - k \bm \epsilon_2 + k \bm \epsilon_3}.
\end{equation*}
One can verify that
\begin{multline*}
A_{k, \bm m}(u) = (-1)^k \kappa_q^{2 k} \, q^{2 k \lambda_2 - 2 k m_3 - k^2 - 2 k} \frac{[m_2]_q!}{[m_2-k]_q!} (1 - q^{2 \lambda_1 - 2 m_1 - 2 m_2 - 2} u) u^k \\*
\times \bigg[ \prod_{i = 0}^k (1 - q^{2 \lambda_1 - 2 m_2 + 2 i} u) (1 - q^{2 \lambda_2 - 2 m_3 - 2 i - 2} u) \bigg]^{-1}.
\end{multline*}
Using this relation, we obtain that
\begin{align*}
\phi^+_{2, \, \bm m}(u) v_{\bm m} & = q^{\lambda_2 - \lambda_3 + m_1 - m_2 - 2 m_3} (1 - q^{2 \lambda_1 - 2 m_1 - 2 m_2 + 1} u) \\* & \times (1 - q^{2 \lambda_1 + 3} u) (1 - q^{2 \lambda_2 + 1} u) (1 - q^{2 \lambda_3 - 1} u) \\ & \times \sum_{k = 0}^{m_2} (-1)^k \kappa_q^{2 k} q^{2 k \lambda_2 - 2 k m_3 - k^2} [k + 1]_q \prod_{i = 1}^k [m_2 -i + 1]_q \, u^k \\ & \times \bigg[ \prod_{i = 0}^{k + 1} (1 - q^{2 \lambda_1 - 2 m_2 + 2 i + 1} u) (1 - q^{2 \lambda_2 - 2 m_3 - 2 i + 1} u) \bigg]^{-1} \, v_{\bm m + k \bm \epsilon_1 - k \bm \epsilon_2 + k \bm \epsilon_3},
\end{align*}
and that for the vectors $w_{\bm m}$ defined by equation (\ref{wm}), where $C_{k, \, \bm m}$ is given by equation (\ref{ckm}), we have
\begin{equation*}
\phi^+_2(u) w_{\bm m} = \Psi^+_{2, \, \bm m}(u) w_{\bm m}
\end{equation*}
with
\begin{multline}
\Psi^+_{2, \, \bm m}(u)
= q^{\lambda_2 - \lambda_3 + m_1 - m_2 - 2 m_3} \, \frac{1 - q^{2 \lambda_1 - 2 m_1 - 2 m_2 + 1} u}{1 - q^{2 \lambda_1 - 2 m_2 + 1} u} \, \frac{1 - q^{2 \lambda_1 + 3} u}{1 - q^{2 \lambda_1 - 2 m_2 + 3} u} \\*
\times \frac{1 - q^{2 \lambda_2 + 1} u}{1 - q^{2 \lambda_2 - 2 m_3 + 1} u} \, \frac{1 - q^{2 \lambda_3 - 1} u}{1 - q^{2 \lambda_2 - 2 m_3 - 1} u}. \label{psip2m}
\end{multline}

Now we proceed to the case of $\phi^-_1(u)$ and $\phi^-_2(u)$. Introducing the generating functions
\begin{equation*}
\bbF'_{\delta, \, \alpha_i}(u^{-1}) = \sum_{n = 1}^\infty \varepsilon(f'_{n \delta, \, \alpha_i}) u^{- n},
\end{equation*}
we obtain
\begin{align}
& \varepsilon(\phi^-_1(u^{-1})) = q^{- K_1 + K_2} \big(1 + \kappa_q \, \bbF'_{\delta, \alpha_1}(- u^{-1})\big), \label{fmpo} \\
& \varepsilon(\phi^-_2(u^{-1})) = q^{- K_2 + K_3} \big(1 + \kappa_q \, \bbF'_{\delta, \alpha_2}(u^{-1})\big). \label{fmpt}
\end{align}
Following the method of the paper \cite{Raz13}, we see that if we define
\begin{align*}
& \bbO'_{11}(u^{-1}) = 1 - u^{-1} q^{- 2 K_1}, && \bbO'_{12} = \kappa_q \, F_1, && \bbO'_{13} = \kappa_q \, F_3, \\
& \bbO'_{21} = \kappa_q \, q \,  \, q^{- K_2 - K_1} \, E_1, && \bbO'_{22}(u^{-1}) = 1 - u^{-1} q^{- 2 K_2}, && \bbO'_{23} = \kappa_q \, F_2, \\*
& \bbO'_{31} = \kappa_q \, q  \, q^{- K_3 - K_1} \, E_3, && \bbO'_{32} = \kappa_q \, q \, q^{- K_3 - K_2} \, E_2,  && \bbO'_{33}(u^{-1}) = 1 - u^{-1} q^{- 2 K_3}
\end{align*}
and the quantities with two and three primes by relations (\ref{mpa})--(\ref{mpb}), where $\bbN$ is changed to $\bbO$ and $u$ to $u^{-1}$, we come to the relations
\begin{align}
& 1 + \kappa_q \, \bbF'_{\delta, \, \alpha_1}(u^{-1}) = \bbO''_{22}(-q^{-2} u^{-1}) \, \bbO'^{-1}_{11}(-q^{-2} u^{-1}) , \label{opfka} \\
& 1 + \kappa_q \, \bbF'_{\delta, \, \alpha_2}(u^{-1}) = \bbO'''_{33}(q^{-3} u^{-1}) \, \bbO''^{-1}_{22}(q^{-3} u^{-1}). \label{opfkb}
\end{align}
Comparing these relations with (\ref{fmpo}) and (\ref{fmpt}), we see that
\begin{align}
& \varepsilon(\phi^-_1(u^{-1})) = q^{- K_1 + K_2} \, \bbO''_{22}(q^{-2} u^{-1}) \, \bbO'^{-1}_{11}(q^{-2} u^{-1}),\\
& \varepsilon(\phi^-_2(u^{-1})) = q^{- K_2 + K_3} \, \bbO'''_{33}(q^{-3} u^{-1}) \, \bbO''^{-1}_{22}(q^{-3} u^{-1}).
\end{align} 

Using the equation
\begin{multline*}
\bbO''_{2 2}(u^{-1}) v_{\bm m} =  \frac{1}{(1 - q^{- 2 \lambda_1 + 2 m_1 + 2 m_2 + 2} u^{-1})} \\*
\times \Big[ (1 - q^{- 2 \lambda_1 + 2 m_2} u^{-1})  (1 - q^{- 2 \lambda_2 + 2 m_3 + 2} u^{-1}) v_{\bm m} \\*
+ \kappa_q^2 \, q^{- 2 \lambda_1 + 2 m_2 - 1} [m_2]_q \, u^{-1} v_{\bm m + \bm \epsilon_1 - \bm \epsilon_2 + \bm \epsilon_3} \Big],
\end{multline*}
we come to the relation
\begin{multline*}
\phi^-_1(u^{-1}) v_{\bm m} =  \frac{q^{- \lambda_1 + \lambda_2 + 2 m_1 + m_2 - m_3}}{(1 - q^{- 2 \lambda_1 + 2 m_1 + 2 m_2 - 2} u^{-1}) (1 - q^{- 2 \lambda_1 + 2 m_1 + 2 m_2} u^{-1})} \\
\times \Big[  (1 - q^{- 2 \lambda_1 + 2 m_2 - 2} u^{-1})  (1 - q^{- 2 \lambda_2 + 2 m_3} u^{-1}) v_{\bm m} \\
+ \kappa_q^2 \, q^{- 2 \lambda_1 + 2 m_2 - 3} [m_2]_q \, u^{-1} v_{\bm m + \bm \epsilon_1 - \bm \epsilon_2 + \bm \epsilon_3} \Big].
\end{multline*}
Now one can verify that for the vectors $w_{\bm m}$ defined by equation (\ref{wm}), where $C_{k, \, \bm m}$ is given by equation (\ref{ckm}), we have
\begin{equation*}
\phi^-_1(u^{-1}) w_{\bm m} = \Psi^-_{1, \, \bm m}(u^{-1}) w_{\bm m},
\end{equation*}
with
\begin{multline}
\Psi^-_{1, \, \bm m}(u^{-1}) \\*
= q^{- \lambda_1 + \lambda_2 + 2 m_1 + m_2 - m_3} \, \frac{1 - q^{- 2 \lambda_1 + 2 m_2 - 2} u^{-1}}{1 - q^{- 2 \lambda_1 + 2 m_1 + 2 m_2 - 2} u^{-1}} \, \frac{1 - q^{- 2 \lambda_2 + 2 m_3} u^{-1}}{1 - q^{- 2 \lambda_1 + 2 m_1 + 2 m_2} u^{-1}}. \label{psim1m}
\end{multline}
To analyze the case of $\phi^-_2(u^{-1})$ we first determine that
\begin{equation*}
\bbO''^{-1}_{22}(u^{-1}) v_{\bm m} = \sum_{k = 0}^{m_2} B_{k, \, \bm m}(u^{-1}) \, v_{\bm m + k \bm \epsilon_1 - k \bm \epsilon_2 + k \bm \epsilon_3},
\end{equation*}
where
\begin{multline*}
B_{k, \, \bm m}(u^{-1}) = (-1)^k \kappa_q^{2 k} \, q^{- 2 k \lambda_1 +2 k m_2 - k^2} \frac{[m_2]_q!}{[m_2-k]_q!} \\ \times (1 - q^{- 2 \lambda_1 + 2 m_1 + 2 m_2 + 2} u^{-1}) \, u^{- k} \\ \times \bigg[ \prod_{i = 1}^k (1 - q^{- 2 \lambda_1 + 2 m_2 - 2 i)} u^{-1}) (1 - q^{- 2 \lambda_2 + 2 m_3 + 2 i + 2} u^{-1}) \bigg]^{-1}.
\end{multline*}
This allows us to obtain the equation
\begin{align*}
\phi^-_2(u^{-1}) v_m & = q^{- \lambda_2 + \lambda_3 - m_1 + m_2 - 2 m_3} (1 - q^{- 2 \lambda_1 + 2 m_1 + 2 m_2 - 1} u^{-1}) \\* & \times (1 - q^{- 2 \lambda_1 - 3} u^{-1}) (1 - q^{- 2 \lambda_2 - 1} u^{-1}) (1 - q^{- 2 \lambda_3 + 1} u^{-1}) \\* & \times \sum_{k = 0}^{m_2} (-1)^k \kappa_q^{2 k} q^{- 2 k \lambda_1 + 2 k m_2 - k^2 - 2 k} [k + 1]_q \prod_{i = 1}^k [m_2 -i + 1]_q \, u^{-k} \\ & \times \bigg[ \prod_{i = 0}^{k + 1} (1 - q^{- 2 \lambda_1 + 2 m_2 - 2 i - 1} u^{-1}) (1 - q^{- 2 \lambda_2 + 2 m_3 + 2 i - 1} u^{-1}) \bigg]^{-1} \\ & \hspace{25.em} \times v_{m + k \epsilon_1 - k \epsilon_2 + k \epsilon_3}.
\end{align*}
which helps us to verify that for the vectors $w_{\bm m}$ defined by equation (\ref{wm}), where $C_{k, \, \bm m}$ is given by equation (\ref{ckm}), we have
\begin{equation*}
\phi^-_2(u^{-1}) w_{\bm m} = \Psi^-_{2, \, \bm m}(u^{-1}) w_{\bm m},
\end{equation*}
with
\begin{multline}
\Psi^-_{2, \, \bm m}(u^{-1}) \\*
= q^{- \lambda_2 + \lambda_3 - m_1 + m_2 + 2 m_3} \, \frac{1 - q^{- 2 \lambda_1 + 2 m_1 + 2 m_2 - 1} u^{-1}}{1 - q^{- 2 \lambda_1 + 2 m_2 - 1} u^{-1}} \, \frac{1 - q^{- 2 \lambda_1 - 3} u^{-1}}{1 - q^{- 2 \lambda_1 + 2 m_2 - 3} u^{-1}} \\*
\times \frac{1 - q^{- 2 \lambda_2 - 1} u^{-1}}{1 - q^{- 2 \lambda_2 + 2 m_3 - 1} u^{-1}} \, \frac{1 - q^{- 2 \lambda_3 + 1} u^{-1}}{1 - q^{- 2 \lambda_2 + 2 m_3 + 1} u^{-1}}. \label{psim2m}
\end{multline}

As we noted above there are two evident automorphisms of $\uqgllpo$ defined by (\ref{sigmab})--(\ref{sigmae}) and (\ref{taub}), (\ref{taue}). From the point of view of integrable systems it is interesting to consider the twisting of the representations $\widetilde \varphi^\lambda$ and $\varphi^\lambda$ by the automorphism~$\tau$. Define the representations
\begin{equation*}
\widetilde \ovarphi{}^\lambda = \widetilde \varphi^\lambda \circ \tau, \qquad \ovarphi{}^\lambda = \varphi^\lambda \circ \tau,
\end{equation*}
and denote the corresponding $\uqlsllpo$-modules as $\widetilde{\oV}{}^\lambda$ and $\oV^\lambda$. It follows from the definitions (\ref{cweb})--(\ref{cwee}) that
\begin{align}
& \tau(e_{n \delta, \, \alpha_1}) = e_{n \delta, \, \alpha_2}, \qquad \tau(e_{n \delta, \, \alpha_2}) = e_{n \delta, \, \alpha_1}, \label{ten} \\
& \tau(f_{n \delta, \, \alpha_1}) = f_{n \delta, \, \alpha_2}, \qquad \tau(f_{n \delta, \, \alpha_2}) = f_{n \delta, \, \alpha_1}, \label{tfn}
\end{align}
see also the paper \cite{Raz13}. Now, it is clear that the basis vectors $w_{\bm m}$ of the modules $\widetilde{\oV}{}^\lambda$ and $\oV^\lambda$ defined by (\ref{wm}) are the $\ell$-weight vectors of $\ell$-weights
\begin{equation*}
\obmPsi_{\bm m} = \{\oPsi^+_{i, \, \bm m}(u), \oPsi^-_{i, \, \bm m}(u^{-1})\}_{i = 1, 2},
\end{equation*}
where
\begin{equation*}
\oPsi^+_{i, \, \bm m}(u) = \Psi^+_{3 - i, \, \bm m}(-u), \qquad \oPsi^-_{i, \, \bm m}(u^{-1}) = \Psi^+_{3 - i, \, \bm m}(-u^{-1}).
\end{equation*}
Here the functions $\Psi^+_{1, \, \bm m}(u)$, $\Psi^+_{2, \bm m}(u)$, $\Psi^-_{1, \, \bm m}(u^{-1})$ and $\Psi^-_{2, \bm m}(u^{-1})$ are given by (\ref{psip1m}), (\ref{psip2m}), (\ref{psim1m}) and (\ref{psim2m}) respectively.

\section{\texorpdfstring{Highest $\ell$-weight representations of Borel subalgebra}{Highest l-weight representations of Borel subalgebra}} \label{s:rbs}

\subsection{General information}

There are two Borel subalgebras of the quantum loop algebra $\uqlg$. In fact, these are the subalgebras whose representations are needed for applications in the theory of integrable systems. In terms of the Drinfeld--Jimbo generators the Borel subalgebras are defined as follows. The Borel subalgebra $\uqbp$ is the subalgebra generated by $e_i$, $i \in \widehat I$, and $q^x$, $x \in \tgothh$, and the Borel subalgebra $\uqbm$ is the subalgebra generated by $f_i$, $i \in \widehat I$, and $q^x$, $x \in \tgothh$. It is clear that these are Hopf subalgebras of $\uqlg$. For a general $\gothg$ there is no such a simple description of $\uqbp$ and $\uqbm$ in terms of the Drinfeld generators. However, it follows from (\ref{ksipin})--(\ref{chiin}) that the Borel subalgebra $\uqbp$ contains the Drinfeld generators $\xi^+_{i, n}$, $\xi^-_{i, m}$, $\chi_{i, m}$ with $i \in I$, $n \geq 0$ and $m > 0$, while the Borel subalgebra $\uqbm$ contains the Drinfeld generators $\xi^-_{i, n}$, $\xi^+_{i, m}$, $\chi_{i, m}$ with $i \in I$, $n \leq 0$ and $m < 0$. The two Borel subalgebras are related by the quantum Chevalley involution. Therefore, we restrict ourselves by the consideration of the subalgebra $\uqbp$.

The definitions of the {\em category $\calO$\/}, a {\em highest weight $\uqbp$-module\/} and the related notions are the same as for the case of $\uqlg$-modules. However, now an {\em $\ell$-weight\/} $\bm \Psi$ is defined as a set
\begin{equation*}
\bm \Psi = \{ \Psi^+_{i, \, n} \in \bbC \mid i \in I, \, n \in \bbZ_+ \}
\end{equation*}
such that $\Psi^+_{i,\, 0} \ne 0$. For any $\uqbp$-module in the category $\calO$ we have the {\em $\ell$-weight decomposition\/}
\begin{equation*}
V = \bigoplus_{\bm \Psi} V_{\bm \Psi},
\end{equation*}
where $V_{\bm \Psi}$ is a subspace of $V$ such that for any $v$ in $V_{\bm \Psi}$ there is $p \in \bbN$ such that
\begin{equation*}
(\phi^+_{i, \, n} - \Psi^+_{i, \, n})^p v = 0
\end{equation*}
for all $i \in I$ and $n \in \bbZ_+$. Similarly as in the case of $\uqlg$-modules, the space $V_{\bm \Psi}$ is called the {\em $\ell$-weight space\/} of $\ell$-weight $\bm \Psi$, and we say that $\bm \Psi$ is an {\em $\ell$-weight\/} of $V$ if $V_{\bm \Psi} \ne \{0\}$. A nonzero element $v \in V_{\bm \Psi}$ such that
\begin{equation*}
\phi^+_{i, \, n} v = \Psi^+_{i, \, n} v
\end{equation*}
for all $i \in I$ and $n \in \bbZ_+$ is said to be an {\em $\ell$-weight vector\/} of $\ell$-weight $\bm \Psi$. As in the case of $\uqlg$-modules, every nontrivial $\ell$-space contains an $\ell$-weight vector.

A $\uqbp$-module $V$ in the category $\calO$ is called a {\em highest $\ell$-weight module\/} with {\em highest $\ell$-weight\/} $\bm \Psi$ if there exists an $\ell$-weight vector $v \in V$ of $\ell$-weight $\bm \Psi$ such that
\begin{equation*}
\xi^+_{i, \, n} v = 0
\end{equation*}
for all $i \in I$ and $n \in \bbZ_+$, and
\begin{equation*}
V = \uqbp v.
\end{equation*}
As in the case of $\uqlg$-modules, the vector with the above properties is unique up to a scalar factor. We again call it the {\em highest $\ell$-weight vector\/} of $V$.

For a given $\ell$-weight $\bm \Psi$ we define the generating function $\Psi^+(u)$ as 
\begin{equation*}
\Psi^+_i(u) = \sum_{n \in \bbZ_+} \Psi^+_{i, \, n} u^n,
\end{equation*}
and, when it is convenient, identify $\bm \Psi$ with the set $\{\Psi^+_i(u)\}_{i \in I}$. An $\ell$-weight $\bm \Psi$ of a $\uqbp$-module is called {\em rational\/} if for some non-negative integers $p_i$, $q_i$, $i \in I$, and complex numbers $a_{i r}$, $b_{i s}$, $i \in I$, $0 \le r \le p_i$, $0 \le s \le q_i$, one has
\begin{equation*}
\Psi^+_i(u) = \frac{a_{i p_i} u^{p_i} + a_{i, \, p_i - 1} u^{p_i - 1} + \cdots + a_{i 0}}{b_{i q_i} u^{q_i} + b_{i, \, q_i - 1} u^{q_i - 1} + \cdots + b_{i 0}}.
\end{equation*}
Here the numbers $a_{i 0}$, $b_{i 0}$ must be nonzero.

As in the case of $\uqlg$-modules, one can show that for any rational $\ell$-weight $\bm \Psi$ there is an irreducible highest $\ell$-weight $\uqbp$-module $L(\bm \Psi)$ with highest $\ell$-weight $\bm \Psi$  which is unique up to an isomorphism, and any irreducible $\uqbp$-module in the category $\calO$ is a highest $\ell$-weight module with a rational highest $\ell$-weight. Here again all $\ell$-weights of a $\uqbp$-module in the category $\calO$ are rational. For any rational $\ell$-weights $\bm \Psi$ and $\bm \Psi'$ the submodule of $L(\bm \Psi) \otimes L(\bm \Psi')$ generated by the tensor product of the highest $\ell$-weight vectors is a highest $\ell$-weight module with highest $\ell$-weight $\bm \Psi \bm \Psi'$. In particular, $L(\bm \Psi \bm \Psi')$ is a subquotient of $L(\bm \Psi) \otimes L(\bm \Psi')$.

The {\em prefundamental representations\/} are the highest $\ell$-weight representations with highest $\ell$-weights determined by the relations
\begin{equation*}
\Psi^+_i(u) = ( \underbracket[.2pt]{1, \, \ldots, \, 1}_{i - 1}, \, (1 - a u)^{\pm 1}, \, \underbracket[.2pt]{1, \, \ldots, \, 1}_{l - i} ), \qquad i \in I, \quad a \in \bbC^\times.
\end{equation*}
The corresponding $\uqbp$-modules are denoted by $L^\pm_{i, \, a}$. For any $\xi \in \gothh^*$ the one dimensional representation with the highest $\ell$-weight defined by the relation
\begin{equation*}
\Psi^+_i(u) = q^{\langle \xi, \, h_i \rangle}, \qquad i \in I
\end{equation*}
is also included into the class of the prefundamental representations. The corresponding $\uqbp$-module is denoted by $L_\xi$.

For any $\uqbp$-module $V$ and an element $\xi \in \tgothh^*$ such that $\langle \xi, \, c \rangle = 0$, we define a {\em shifted $\uqbp$-module\/} $V[\xi]$ shifting the action of the generators $q^x$. Namely, if $\varphi$ is the representation of $\uqbp$ corresponding to the module $V$ and $\varphi[\xi]$ is the representation corresponding to the module $V[\xi]$, then
\begin{equation*}
\varphi[\xi](e_i) = \varphi(e_i), \quad i \in I, \qquad \varphi[\xi](q^x) = q^{\langle \xi, \, x \rangle} \varphi(q^x), \quad x \in \tgothh.
\end{equation*}
Remind that an element $\xi \in \tgothh^*$ satisfying the relation $\langle \xi, \, c \rangle = 0$ can be naturally identified with an element of $\gothh^*$. It is clear that the module $V[\xi]$ is isomorphic to $V \otimes L_\xi$.

One can show that any $\uqbp$-module in the category $\calO$ is a subquotient of a tensor product of prefundamental representations.

\subsection{\texorpdfstring{$q$-oscillators}{q-oscillators}}

To obtain a representation of a Borel subalgebra one can simply take the restriction of a representation of the full quantum loop algebra to this subalgebra. However,  for the theory of integrable systems more representations are needed. Here one constructs necessary representations first defining a homomorphism of a Borel subalgebra to the $q$-oscillator algebra or to the tensor product of several copies of this algebra. Then one uses the appropriate representations of the $q$-oscillator algebras and comes to the desirable representation of the Borel subalgebra. In this section we give the definition of the $q$-oscillator algebra and describe its important representations.

Let $\hbar$ be a non-zero complex number and $q = \exp \hbar$.\footnote{We again assume that $q$ is not a root of unity.} The $q$-oscillator algebra $\Osc_q$ is a unital associative $\bbC$-algebra with generators $b^\dagger$, $b$, $q^{\nu N}$, $\nu \in \bbC$, and relations
\begin{gather*}
q^0 = 1, \qquad q^{\nu_1 N} q^{\nu_2 N} = q^{(\nu_1 + \nu_2)N}, \\
q^{\nu N} b^\dagger q^{-\nu N} = q^\nu b^\dagger, \qquad q^{\nu N} b q^{-\nu N} = q^{-\nu} b, \\
b^\dagger b = \frac{q^N - q^{- N}}{q - q^{-1}}, \qquad b b^\dagger = \frac{q q^N - q^{-1} q^{- N}}{q - q^{-1}}.
\end{gather*}
Two representations of $\Osc_q$ are interesting for us. First, let $W^{\scriptscriptstyle +}$ be the free vector space generated by the set $\{ v_0, v_1, \ldots \}$. One can show that the relations
\begin{gather}
q^{\nu N} v_m = q^{\nu m} v_m, \label{vpm} \\*
b^\dagger v_m = v_{m + 1}, \qquad b \, v_m = [m]_q v_{m - 1},
\end{gather}
where we assume that $v_{-1} = 0$, endow $W^{\scriptscriptstyle +}$ with the structure of an $\Osc_q$-module. We denote the corresponding representation of the algebra $\Osc_q$ by $\chi^{\scriptscriptstyle +}$. Further, let $W^{\scriptscriptstyle -}$ be the free vector space generated again by the set $\{ v_0, v_1, \ldots \}$. The relations
\begin{gather}
q^{\nu N} v_m = q^{- \nu (m + 1)} v_m, \label{vmm} \\
b \, v_m = v_{m + 1}, \qquad b^\dagger v_m = - [m]_q v_{m - 1},
\end{gather}
where we again assume that $v_{-1} = 0$, endow the vector space $W^{\scriptscriptstyle -}$ with the structure of an $\Osc_q$-module. We denote the corresponding representation of $\Osc_q$ by $\chi^{\scriptscriptstyle -}$.

\subsection{\texorpdfstring{Case of $\gothg = \slii$}{Case of g = sl2}}

\subsubsection{Definition of representations}

One can show that the mapping $\rho: \uqbp \to \Osc_q$ defined by the relations
\begin{align*}
& \rho(q^{\nu h_0}) = q^{2 \nu N}, && \rho(q^{\nu h_1}) = q^{- 2\nu N}, \\*
& \rho(e_0) = b^\dagger, && \rho(e_1) = - \kappa_q^{-1} b \, q^N 
\end{align*}
is a homomorphism from the Borel subalgebra $\uqbp$ to the algebra $\Osc_q$. Using this homomorphism we define two representations of $\uqbp$:
\begin{equation*}
\theta_1 = \chi^- \circ \rho \circ \sigma^{- 1}, \qquad \theta_2 = \chi^+ \circ \rho.
\end{equation*}
Here the representations of $\Osc_q$ are chosen so to get highest $\ell$-weight representations. Let us find all $\ell$-weights for these representations. We give only a few intermediate formulas. 

\subsubsection{\texorpdfstring{Representation $\theta_1$}{Representation theta1}}

The vectors
\begin{equation*}
v_m = b^m v_0, \qquad m \in \bbZ_+,
\end{equation*}
form a basis in the representation space. Direct calculations give
\begin{equation*}
\theta_1(e'_{n \delta}) = \chi^- \big(\kappa_q^{-1} (-1)^{n - 1} q^{2 n} ([n + 1]_q - q^{-1} [n]_q q^{- 2 N_1}) q^{2 n N_1}\big),
\end{equation*}
see the papers \cite{BooGoeKluNirRaz10, BooGoeKluNirRaz14a}. It follows from this equation that
\begin{equation*}
1 - \kappa_q \, \bbE'_\delta(u) = \chi^- \big((1 + q u)(1 + q^3 q^{2 N_1} u)^{-1} (1 + q q^{2 N_1} u)^{-1} \big),
\end{equation*}
where the generating function is defined as\footnote{Below we use similar natural relations to define necessary generating functions not writing them explicitly.}
\begin{equation*}
\bbE'_\delta(u) = \sum_{n = 1}^\infty \theta_1 (e_{n \delta}) u^n.
\end{equation*}
 Now, using (\ref{phipiu}) and (\ref{vmm}), we obtain
\begin{equation*}
\phi^+(u) \, v_m = \Psi^+_{m, \, 1}(u) \, v_m = q^{- 2 m - 2} \frac{1 - q u}{(1 - q^{- 2 m + 1} u)(1 - q^{- 2 m - 1} u)} \, v_m.
\end{equation*}

\subsubsection{\texorpdfstring{Representation $\theta_2$}{Representation theta2}}

In this case we use the basis formed by the vectors
\begin{equation*}
v_m = (b^\dagger)^m v_0, \qquad m \in \bbZ_+.
\end{equation*}
After some simple calculations we obtain
\begin{equation*}
\theta_2(e'_{\delta}) = \kappa_q^{-1} q, \qquad \theta_2(e'_{n \delta}) = 0, \quad n > 1,
\end{equation*}
see the papers \cite{BooGoeKluNirRaz10, BooGoeKluNirRaz14a, NirRaz14}. This gives
\begin{equation*}
1 - \kappa_q \bbE'_\delta(u) = 1 + q u,
\end{equation*}
and, again taking into account (\ref{phipiu}), we come to the equation
\begin{equation*}
\phi^+(u) \, v_m = \Psi^+_{m, \, 2}(u) \, v_m = (1 - q u) \, v_m.
\end{equation*}

\subsection{\texorpdfstring{Case of $\gothg = \sliii$}{Case of g = sl3}} \label{s:bsl3}

Consider the algebra $\Osc_q \otimes \Osc_q$. As is usual, define
\begin{gather*}
b^{}_1 = b \otimes 1, \qquad b^\dagger_1 = b^\dagger \otimes 1, \qquad b^{}_2 = 1 \otimes b, \qquad b^\dagger_2 = 1 \otimes b^\dagger, \\
q^{\nu_1 N_1 + \nu_2 N_2} = q^{\nu_1 N} \otimes q^{\nu_2 N}
\end{gather*}
The homomorphism in question from $\uqbp$ to $\Osc_q \otimes \Osc_q$ is defined by the relations
\begin{align*}
& \rho(q^{\nu h_0}) = q^{\nu(2 N_1 + N_2)}, && \rho(q^{\nu h_1}) = q^{\nu(- N_1 + N_2)}, &&
\rho(q^{\nu h_2}) = q^{\nu(- N_1 - 2 N_2)}, \\*
& \rho(e_0) = b^\dagger_1 q^{N_2}, && \rho(e_1) = - q^{-1} b_1^{\mathstrut} b_2^\dagger q^{N_1 - N_2}, && \rho(e_2) = - \kappa_q^{-1} b_2^{\mathstrut} q^{N_2}.
\end{align*}
Now we define six representations of $\uqbp$:
\begin{align}
& \theta_1 = (\chi^- \otimes \chi^-) \circ \rho \circ \sigma^{- 1}, && \otheta_1 = (\chi^+ \otimes \chi^+) \circ \rho \circ \tau, \label{tib} \\
& \theta_2 = (\chi^- \otimes \chi^+) \circ \rho \circ \sigma^{- 2}, && \otheta_2 = (\chi^- \otimes \chi^+) \circ \rho \circ \sigma^{-2} \circ \tau, \\
& \theta_3 = (\chi^+ \otimes \chi^+) \circ \rho, && \otheta_3 = (\chi^- \otimes \chi^-) \circ \rho \circ \sigma^{-1} \circ \tau. \label{tie}
\end{align}
The representations for $q$-oscillators are again chosen so to get highest $\ell$-weight representations. The calculation necessary to find $\ell$-weights for these representations are more complicated. Nevertheless, we again give only a few intermediate formulas, referring to our previous papers. 

\subsubsection{\texorpdfstring{Representation $\theta_1$}{Representation theta1}}

For this case we use the basis of the representation space for\-med by the vectors
\begin{equation}
v_{\bm m} = b_1^{m_1} b_2^{m_2} \, v_{\bm 0}, \label{vmmm}
\end{equation}
where $m_1, m_2 \in \bbZ_+$, $\bm m = (m_1, \, m_2)$ and $v_{\bm 0} = v_{(0, \, 0)} = v_0 \otimes v_0$. Direct calculations give
\begin{align*}
& \theta_1(e'_{n \delta, \, \alpha_1}) = (\chi^- \otimes \chi^-) \big( \kappa_q^{- 1} (-1)^{n - 1} q^{3 n}([n + 1]_q - q^{- 1} [n]_q q^{- 2 N_1}) q^{2 n N_1 + 2 n N_2} \big), \\
& \theta_1(e'_{n \delta, \, \alpha_2}) = (\chi^- \otimes \chi^-) \big( - \kappa_q^{- 1} q^{2 n}([n + 1]_q - q^{- 1} [n]_q q^{- 2 N_2} \\
& \hspace{18.em} {} - q [n]_q q^{2 N_1} + [n - 1]_q q^{2 N_1 - 2 N_2} ) q^{2 n N_2} \big),
\end{align*}
see the paper \cite{BooGoeKluNirRaz10} for similar calculations.\footnote{Note that in the paper \cite{BooGoeKluNirRaz10} another definition of $q$-operators is used. However, it is not difficult to adopt the calculations given there to our case.} Using these relation we come to the following expressions for the generating functions $\bbE'_{\delta, \, \alpha_1}(u)$ and $\bbE'_{\delta, \, \alpha_2}(u)$:
\begin{align*}
& 1 - \kappa_q \, \bbE'_{\delta, \, \alpha_1}(u) = (\chi^- \otimes \chi^-) \big( (1 + q^2 q^{2 N_2} u) \\
& \hspace{16.em} {} \times (1 + q^4 q^{2 N_1 + 2 N_2} u)^{-1} (1 + q^2 q^{2N_1 + 2 N_2} u)^{-1} \big), \\
& 1 - \kappa_q \, \bbE'_{\delta, \, \alpha_2}(u) = (\chi^- \otimes \chi^-) \big( (1 - q u) (1 - q^3 q^{2 N_1 + 2 N_2} u) \\
& \hspace{20.em} {} \times (1 - q^3 q^{2 N_2} u)^{-1} (1 - q q^{2 N_2} u)^{-1} \big).
\end{align*}
Now, using (\ref{phipiu}) and (\ref{vmm}), we obtain
\begin{align}
& \phi^+_1(u) v_{\bm m} = \Psi^+_{1, \, \bm m, \, 1}(u) \, v_{\bm m} = q^{- 2 m_1 - m_2 - 3} \frac{1 - q^{- 2 m_2} u}{(1 - q^{- 2 m_1 - 2 m_2} u) (1 - q^{2 m_1 - 2 m_2 - 2} u)} \, v_{\bm m}, \label{psip1m1} \\
& \phi^+_2(u) v_{\bm m} = \Psi^+_{2, \, \bm m, \, 1}(u) \, v_{\bm m} = q^{m_1 - m_2} \frac{(1 - q u) (1 - q^{-2 m_1 - 2 m_2 - 1} u)}{(1 - q^{- 2 m_2 + 1} u) (1 - q^{- 2 m_2 - 1} u)} \, v_{\bm m}. \label{psip2m1}
\end{align}

\subsubsection{\texorpdfstring{Representation $\theta_2$}{Representation theta2}}

Here the natural basis in the representation space is formed by the vectors
\begin{equation}
v_{\bm m} = b_1^{m_1} (b_2^\dagger)^{m_2} \, v_{\bm 0}. \label{vmmp}
\end{equation}
Similarly as in the previous case one obtains that
\begin{align*}
& \theta_2(e'_{1 \delta, \, \alpha_1}) = (\chi^- \otimes \chi^+) \big( - \kappa_q^{-1} q^2 q^{2 N_1} \big), \qquad \theta_2(e'_{n \delta, \, \alpha_1}) = 0, \quad n > 1, \\
& \theta_2(e'_{n \delta, \, \alpha_2}) = (\chi^- \otimes \chi^+) \big( - \kappa_q^{-1} q^{2 n} ([n + 1]_q - q^{-1} [n]_q q^{- 2 N_1}) q^{2 n N_1} \big),
\end{align*}
and come to the equations
\begin{align*}
& 1 - \kappa_q \, \bbE'_{\delta, \, \alpha_1}(u) = (\chi^- \otimes \chi^+) \big( 1 + q^2 q^{2 N_1} u \big), \\
& 1 - \kappa_q \, \bbE'_{\delta, \, \alpha_2}(u) = (\chi^- \otimes \chi^+) \big( (1 - q u) (1 - q^3 q^{2 N_1} u)^{-1} (1 - q q^{2 N_1} u)^{-1} \big).
\end{align*}
Using (\ref{phimiu}), (\ref{vpm}) and (\ref{vmm}), we determine that
\begin{align}
& \phi^+_1(u) v_{\bm m} = \Psi^+_{1, \, \bm m, \, 2}(u) \, v_{\bm m} = q^{m_1 - 2 m_2 + 1} (1 - q^{- 2 m_1} u) \, v_{\bm m}, \label{psip1m2} \\
& \phi^+_2(u) v_{\bm m} = \Psi^+_{2, \, \bm m, \, 2}(u) \, v_{\bm m} = q^{- 2 m_1 + m_2 - 2} \frac{1 - q u}{(1 - q^{- 2 m_1 + 1} u)(1 - q^{- 2 m_1 - 1} u)} \, v_{\bm m}. \label{psip2m2}
\end{align}

\subsubsection{\texorpdfstring{Representation $\theta_3$}{Representation theta3}}

In accordance with the definition of the representation $\theta_3$ we introduce the basis in the representation space formed by the vectors
\begin{equation}
v_{\bm m} = (b_1^\dagger)^{m_1} (b_2^\dagger)^{m_2} \, v_{\bm 0}. \label{vmpp}
\end{equation}
Here the necessary calculations are very simple and one obtains
\begin{equation*}
\theta_3(e'_{n \delta, \, \alpha_1}) = 0, \qquad \theta_3(e'_{1 \delta, \, \alpha_2}) = \kappa_q^{-1} q, \qquad \theta_3(e'_{n \delta, \, \alpha_1}) = 0, \quad n > 1.
\end{equation*}
Hence, we see that
\begin{align*}
1 - \kappa_q \, \bbE'_{\delta, \, \alpha_1}(u) = 1, \qquad 1 - \kappa_q \, \bbE'_{\delta, \, \alpha_1}(u) = 1 - q u,
\end{align*}
and, using (\ref{phipiu}) and (\ref{vpm}), come to the final result
\begin{align}
& \phi^+_1(u) \, v_{\bm m} = \Psi^+_{1, \, \bm m, \, 3}(u) \, v_{\bm m} = q^{- m_1 + m_2} \, v_{\bm m}, \label{psip1m3} \\
& \phi^+_2(u) \, v_{\bm m} = \Psi^+_{2, \, \bm m, \, 3}(u) \, v_{\bm m} = q^{- m_1 - 2 m_2} (1 - q u) \, v_{\bm m}. \label{psip2m3}
\end{align}

\subsubsection{\texorpdfstring{Representations $\otheta_1$, $\otheta_2$ and $\otheta_3$}{Representations theta1, theta2 and theta3}}

Taking into account relations (\ref{ten}) and the definition (\ref{tib})--(\ref{tie}) of the considered representations, we conclude that the corresponding basis vectors $v_{\bm m}$ defined by (\ref{vmpp}), (\ref{vmmp}), or by (\ref{vmmm}) are the $\ell$-weight vectors of $\ell$-weights
\begin{equation*}
\obmPsi_{\bm m, \, a} = \{\oPsi^+_{i, \, \bm m, \, a}(u)\}_{i = 1, 2}, \qquad a = 1, 2, 3,
\end{equation*}
where
\begin{equation*}
\oPsi^+_{i, \, \bm m, \, a}(u) = \Psi^+_{3 - i, \, \bm m, \, 4 - a}(-u).
\end{equation*}
Here the functions $\Psi^+_{i, \, \bm m, \, a}$ are given by equations (\ref{psip1m3}), (\ref{psip2m3}), (\ref{psip1m2}), (\ref{psip2m2}), (\ref{psip1m1}) and (\ref{psip2m1}).

\section{Discussion}

We have obtained the $\ell$-weights and the corresponding $\ell$-weight vectors for representations of quantum loop algebras $\uqlsllpo$ with $l = 1, 2$ obtained via Jimbo's homomorphism, known also as evaluation representations. It appears that the representation space has a basis consisting of $\ell$-weight vectors. This means that the number $p$ in (\ref{pppmo}) is always equal to 1. Then we have found the $\ell$-weights and the $\ell$-weight vectors for the $q$-oscillator representations of Borel subalgebras of the same quantum loop algebras, and again discovered that for all representations the representation space has a basis consisting of $\ell$-weight vectors. We see that some $q$-oscillator representations are shifted prefundamental representations, and any prefundamental representation is presented in a shifted form among the $q$-oscillator representations.

In applications to the theory of quantum integrable systems one associates with a representation of a quantum loop algebra or a family of representations parametrized by the so called {\em spectral parameter}.
The usual way to do this is as follows. Given $\zeta \in \bbC^\times$, we define an automorphism $\Gamma_\zeta$ of $\uqlg$ by its action on the generators as
\begin{equation*}
\Gamma_\zeta(e_i) = \zeta^{s_i} e_i, \qquad \Gamma_\zeta(f_i) = \zeta^{-s_i} f_i, \qquad \Gamma_\zeta(q^x) = q^x,
\end{equation*}
where $s_i$ are arbitrary integers. Then, starting from a representation $\varphi$ of $\uqlsllpo$ we define the  family of representations $\varphi_\zeta$ in question as
\begin{equation*}
\varphi_\zeta = \varphi \circ \Gamma_\zeta.
\end{equation*}
In a similar way, one defines for the Borel subalgebras families of representations parametrized by the spectral parameter.

Let $\varphi$ be a representation of $\uqlg$ and $V$ be the corresponding $\uqlg$-module. We denote by $V_\zeta$ the $\uqlg$-module corresponding to the representation $\varphi_\zeta$. If $V$ is a highest $\ell$-weight module with highest $\ell$-weight determined by the functions $\Psi^+_i(u)$ and $\Psi^-_i(u^{-1})$ then $V_\zeta$ is a highest $\ell$-weight $\uqlg$-module with highest $\ell$-weight determined by the functions $\Psi^+_i(\zeta^s u)$ and $\Psi^-_i(\zeta^{-s} u^{-1})$, where $s = s_0 + s_1 + \cdots + s_l$.

Denote the $\uqbp$-modules corresponding to the representations $\theta_1$, $\theta_2$ and $\theta_3$ defined in (\ref{tib})--(\ref{tie}) as $W_1$, $W_2$ and $W_3$, and consider the $\uqbp$-module $(W_1)_{\zeta_1} \otimes (W_2)_{\zeta_2} \otimes (W_3)_{\zeta_3}$. As follows from results of section \ref{s:bsl3}, the tensor product of the highest $\ell$-weight vectors is an $\ell$-weight vector of $\ell$-weight determined by the functions
\begin{equation*}
\Psi^+_1(u) = q^{-2} \, \frac{1 - \zeta_2^s u}{1 - q^{-2} \zeta_1^s u}, \qquad \Psi^+_2(u) = q^{-2} \, \frac{1 - q \zeta_3^s u}{1 - q^{-1} \zeta_2^s u}.
\end{equation*}
Consider now the restriction of the representation $V^\lambda$ to the Borel subalgebra $\uqbp$. We denote this restriction again by $V^\lambda$. Using results of section \ref{s:fsl3}, we see that the highest $\ell$-weight of the $\uqbp$-module $(\widetilde V^\lambda)_\zeta$ is determined by the functions
\begin{equation*}
\Psi^+_1(u) = q^{\lambda_1 - \lambda_2} \, \frac{1 - q^{2 \lambda_2} \zeta^s u}{1 - q^{2 \lambda_1} \zeta^s u}, \qquad \Psi^+_2(u) = q^{\lambda_2 - \lambda_3} \, \frac{1 - q^{2 \lambda_3 - 1} \zeta^s u}{1 - q^{2 \lambda_2 - 1} \zeta^s u}.
\end{equation*}
It follows that if
\begin{equation*}
\zeta_1 = q^{2 (\lambda_1 + 1)} \zeta, \qquad \zeta_2 = q^{2 \lambda_2} \zeta, \qquad \zeta_3 = q^{2 (\lambda_3 - 1)} \zeta,
\end{equation*}
then the submodule of $(W_1)_{\zeta_1} \otimes (W_2)_{\zeta_2} \otimes (W_3)_{\zeta_3}$ generated by the tensor product of the highest $\ell$-weight vectors of $(W_1)_{\zeta_1}$, $(W_2)_{\zeta_2}$ and $(W_3)_{\zeta_3}$ is isomorphic to the shifted module $(\widetilde V^\lambda)_\zeta[\xi]$, where $\xi$ is determined by the equations
\begin{equation*}
\xi(h_1) = - \lambda_1 + \lambda_2 - 2, \qquad \xi(h_1) = - \lambda_2 + \lambda_3 - 2.
\end{equation*}
This result is in the full agreement with that obtained in the paper \cite{BooGoeKluNirRaz14b} by explicit analysis of the tensor product of the modules. The results of such kind are important for establishing functional relations. We see that they can be obtained by considering $\ell$-weights of the representations.

\subsection*{Acknowledgments} This work was supported in part by the DFG grant KL \hbox{645/10-1}. Kh.S.N. and A.V.R. were supported in part by the RFBR grants \#~13-01-00217 and \#~14-01-91335.

\newcommand{\noopsort}[1]{}
\providecommand{\bysame}{\leavevmode\hbox to3em{\hrulefill}\thinspace}
\providecommand{\href}[2]{#2}
\providecommand{\curlanguage}[1]{%
 \expandafter\ifx\csname #1\endcsname\relax
 \else\csname #1\endcsname\fi}


\begin{thebibliography}{10}

\bibitem{BazLukZam96}
\curlanguage{English}
V.~V. Bazhanov, S.~L. Lukyanov, and A.~B. Zamolodchikov, \emph{Integrable
  structure of conformal field theory, quantum {K}d{V} theory and thermodynamic
  {B}ethe ansatz}, \href{http://dx.doi.org/10.1007/BF02101898}{Commun. Math.
  Phys.} \textbf{177} (1996), 381--398,
  \href{http://arxiv.org/abs/hep-th/9412229}{{\tt arXiv:hep-th/9412229}}.

\bibitem{BazLukZam97}
\curlanguage{English}
V.~V. Bazhanov, S.~L. Lukyanov, and A.~B. Zamolodchikov, \emph{Integrable
  structure of conformal field theory {II}. {Q}-operator and {DDV} equation},
  \href{http://dx.doi.org/10.1007/s002200050240}{Commun. Math. Phys.}
  \textbf{190} (1997), 247--278,
  \href{http://arxiv.org/abs/hep-th/9604044}{{\tt arXiv:hep-th/9604044}}.

\bibitem{BazLukZam99}
\curlanguage{English}
V.~V. Bazhanov, S.~L. Lukyanov, and A.~B. Zamolodchikov, \emph{Integrable
  structure of conformal field theory {III}. {T}he {Y}ang--{B}axter relation},
  \href{http://dx.doi.org/10.1007/s002200050531}{Commun. Math. Phys.}
  \textbf{200} (1999), 297--324,
  \href{http://arxiv.org/abs/hep-th/9805008}{{\tt arXiv:hep-th/9805008}}.

\bibitem{KhoTol92}
\curlanguage{English}
S.~M. Khoroshkin and V.~N. Tolstoy, \emph{The uniqueness theorem for the
  universal {$R$}-matrix}, \href{http://dx.doi.org/10.1007/BF00402899}{Lett.
  Math. Phys.} \textbf{24} (1992), 231--244.

\bibitem{LevSoiStu93}
\curlanguage{English}
S.~Levenderovski\u\i{}, Ya. Soibelman, and V.~Stukopin, \emph{The quantum
  {W}eyl group and the universal quantum {$R$}-matrix for affine {L}ie algebra
  {$A_1^{(1)}$}}, \href{http://dx.doi.org/10.1007/BF00777372}{Lett. Math.
  Phys.} \textbf{27} (1993), 253--264.

\bibitem{ZhaGou94}
\curlanguage{English}
Y.-Z. Zhang and M.~D. Gould, \emph{Quantum affine algebras and universal
  ${R}$-matrix with spectral parameter},
  \href{http://dx.doi.org/10.1007/BF00750144}{Lett. Math. Phys.} \textbf{31}
  (1994), 101--110, \href{http://arxiv.org/abs/hep-th/9307007}{{\tt
  arXiv:hep-th/9307007}}.

\bibitem{BraGouZhaDel94}
\curlanguage{English}
A.~J. Bracken, M.~D. Gould, Y.-Z. Zhang, and G.~W. Delius, \emph{Infinite
  families of gauge-equivalent {$R$}-matrices and gradations of quantized
  affine algebras}, \href{http://dx.doi.org/10.1142/S0217979294001585}{Int. J.
  Mod. Phys. B} \textbf{8} (1994), 3679--3691,
  \href{http://arxiv.org/abs/hep-th/9310183}{{\tt arXiv:hep-th/9310183}}.

\bibitem{BraGouZha95}
\curlanguage{English}
A.~J. Bracken, M.~D. Gould, and Y.-Z. Zhang, \emph{Quantised affine algebras
  and parameter-dependent {$R$}-matrices},
  \href{http://dx.doi.org/10.1017/S0004972700014040}{Bull. Austral. Math. Soc.}
  \textbf{51} (1995), 177--194.

\bibitem{BooGoeKluNirRaz10}
\curlanguage{English}
H.~Boos, F.~G{\"o}hmann, A.~Kl{\"u}mper, Kh.~S. Nirov, and A.~V. Razumov,
  \emph{Exercises with the universal {$R$}-matrix},
  \href{http://dx.doi.org/10.1088/1751-8113/43/41/415208}{J. Phys. A: Math.
  Theor.} \textbf{43} (2010), 415208 (35pp),
  \href{http://arxiv.org/abs/1004.5342}{{\tt arXiv:1004.5342 [math-ph]}}.

\bibitem{BooGoeKluNirRaz11}
\curlanguage{English}
H.~Boos, F.~G{\"o}hmann, A.~Kl{\"u}mper, Kh.~S. Nirov, and A.~V. Razumov,
  \emph{On the universal ${R}$-matrix for the {I}zergin--{K}orepin model},
  \href{http://dx.doi.org/10.1088/1751-8113/44/35/355202}{J. Phys. A: Math.
  Theor.} \textbf{44} (2011), 355202 (25pp),
  \href{http://arxiv.org/abs/1104.5696}{{\tt arXiv:1104.5696 [math-ph]}}.

\bibitem{BazTsu08}
\curlanguage{English}
V.~V. Bazhanov and Z.~Tsuboi, \emph{Baxter's {Q}-operators for supersymmetric
  spin chains}, \href{http://dx.doi.org/10.1016/j.nuclphysb.2008.06.025}{Nucl.
  Phys. B} \textbf{805} (2008), 451--516,
  \href{http://arxiv.org/abs/0805.4274}{{\tt arXiv:0805.4274 [hep-th]}}.

\bibitem{BooGoeKluNirRaz13}
\curlanguage{English}
H.~Boos, F.~G{\"o}hmann, A.~Kl{\"u}mper, Kh.~S. Nirov, and A.~V. Razumov,
  \emph{Universal integrability objects},
  \href{http://dx.doi.org/10.1007/s11232-013-0002-8}{Theor. Math. Phys.}
  \textbf{174} (2013), 21--39, \href{http://arxiv.org/abs/1205.4399}{{\tt
  arXiv:1205.4399 [math-ph]}}.

\bibitem{Raz13}
\curlanguage{English}
A.~V. Razumov, \emph{Monodromy operators for higher rank},
  \href{http://dx.doi.org/10.1088/1751-8113/46/38/385201}{J. Phys. A: Math.
  Theor.} \textbf{46} (2013), 385201 (24pp),
  \href{http://arxiv.org/abs/1211.3590}{{\tt arXiv:1211.3590 [math.QA]}}.

\bibitem{BooGoeKluNirRaz14a}
\curlanguage{English}
H.~Boos, F.~G{\"o}hmann, A.~Kl\"umper, Kh.~S. Nirov, and A.~V. Razumov,
  \emph{Universal ${R}$-matrix and functional relations},
  \href{http://dx.doi.org/10.1142/S0129055X14300052}{Rev. Math. Phys.}
  \textbf{26} (2014), 1430005 (66pp),
  \href{http://arxiv.org/abs/1205.1631}{{\tt arXiv:1205.1631 [math-ph]}}.

\bibitem{BazHibKho02}
\curlanguage{English}
V.~V. Bazhanov, A.~N. Hibberd, and S.~M. Khoroshkin, \emph{Integrable structure
  of {$\mathcal W_3$} conformal field theory, quantum {B}oussinesq theory and
  boundary affine {T}oda theory},
  \href{http://dx.doi.org/10.1016/S0550-3213(01)00595-8}{Nucl. Phys. B}
  \textbf{622} (2002), 475--574,
  \href{http://arxiv.org/abs/hep-th/0105177}{{\tt arXiv:hep-th/0105177}}.

\bibitem{Koj08}
\curlanguage{English}
T.~Kojima, \emph{Baxter's ${Q}$-operator for the ${W}$-algebra ${W_N}$},
  \href{http://dx.doi.org/10.1088/1751-8113/41/35/355206}{J. Phys. A: Math.
  Theor} \textbf{41} (2008), 355206 (16pp),
  \href{http://arxiv.org/abs/0803.3505}{{\tt arXiv:0803.3505 [nlin.SI]}}.

\bibitem{BooGoeKluNirRaz14b}
\curlanguage{English}
H.~Boos, F.~G{\"o}hmann, A.~Kl\"umper, Kh.~S. Nirov, and A.~V. Razumov,
  \emph{Quantum groups and functional relations for higher rank},
  \href{http://dx.doi.org/10.1088/1751-8113/47/27/275201}{J. Phys. A: Math.
  Theor.} \textbf{47} (2014), 275201 (47pp),
  \href{http://arxiv.org/abs/1312.2484}{{\tt arXiv:1312.2484 [math-ph]}}.

\bibitem{NirRaz14}
\curlanguage{English}
Kh.~S. Nirov and A.~V. Razumov, \emph{Quantum groups and functional relations
  for lower rank}, \href{http://arxiv.org/abs/1412.7342}{{\tt arXiv:1412.7342
  [math-ph]}}.

\bibitem{HerJim12}
\curlanguage{English}
D.~Hernandez and Jimbo M, \emph{Asymptotic representations and {D}rinfeld
  rational fractions}, \href{http://dx.doi.org/10.1112/S0010437X12000267}{Comp.
  Math.} \textbf{148} (2012), 1593--1623,
  \href{http://arxiv.org/abs/1104.1891}{{\tt arXiv:1104.1891 [math.QA]}}.

\bibitem{FreHer15}
\curlanguage{English}
E.~Frenkel and D.~Hernandez, \emph{Baxter's relations and spectra of quantum
  integrable models}, Duke Math. J. \textbf{164} (2015), 2407--2460,
  \href{http://arxiv.org/abs/1308.3444}{{\tt arXiv:1308.3444 [math.QA]}}.

\bibitem{MukYou14}
\curlanguage{English}
E.~Mukhin and C.~A.~S. Young, \emph{Affinization of category $\mathcal{O}$ for
  quantum groups},
  \href{http://dx.doi.org/10.1090/S0002-9947-2014-06039-X}{Trans. Amer. Math.
  Soc.} \textbf{366} (2014), 4815--4847,
  \href{http://arxiv.org/abs/1204.2769}{{\tt arXiv:1204.2769 [math.QA]}}.

\bibitem{Kac90}
\curlanguage{English}
V.~Kac, \emph{Infinite-dimensional {L}ie algebras}, Cambridge University Press,
  Cambridge, 1990.

\bibitem{TolKho92}
\curlanguage{English}
V.~N. Tolstoy and S.~M. Khoroshkin, \emph{The universal {$R$}-matrix for
  quantum untwisted affine {L}ie algebras},
  \href{http://dx.doi.org/10.1007/BF01077085}{Funct. Anal. Appl.} \textbf{26}
  (1992), 69--71.

\bibitem{KhoTol93}
\curlanguage{English}
S.~M. Khoroshkin and V.~N. Tolstoy, \emph{On {D}rinfeld's realization of
  quantum affine algebras},
  \href{http://dx.doi.org/10.1016/0393-0440(93)90070-U}{J. Geom. Phys.}
  \textbf{11} (1993), 445--452.

\bibitem{Bec94a}
\curlanguage{English}
J.~Beck, \emph{Convex bases of {PBW} type for quantum affine algebras},
  \href{http://dx.doi.org/10.1007/BF02099742}{Commun. Math. Phys.} \textbf{165}
  (1994), 193--199, \href{http://arxiv.org/abs/hep-th/9407003}{{\tt
  arXiv:hep-th/9407003}}.

\bibitem{AshSmiTol79}
\curlanguage{English}
R.~M. Asherova, Yu.~F. Smirnov, and V.~N. Tolstoy, \emph{Description of a class
  of projection operators for semisimple complex lie algebras},
  \href{http://dx.doi.org/10.1007/BF01140268}{Math. Notes} \textbf{26} (1979),
  499--504.

\bibitem{Tol89}
\curlanguage{English}
V.~N. Tolstoy, \emph{Extremal projections for contragredient {L}ie algebras and
  superalgebras of finite growth},
  \href{http://dx.doi.org/10.1070/RM1989v044n01ABEH002023}{Russian Math.
  Surveys} \textbf{44} (1989), 267--258.

\bibitem{Dri87}
\curlanguage{English}
V.~G. Drinfeld, \emph{Quantum groups}, Proceedings of the International
  Congress of Mathematicians, Berkeley, 1986 (A.~E. Gleason, ed.), vol.~1,
  American Mathematical Society, Providence, 1987, pp.~798--820.

\bibitem{Dri88}
\curlanguage{English}
V.~G. Drinfeld, \emph{A new realization of {Y}angians and quantized affine
  algebras}, Soviet Math. Dokl. \textbf{36} (1988), 212--216.

\bibitem{KhoTol94}
\curlanguage{English}
S.~Khoroshkin and V.~N. Tolstoy, \emph{Twisting of quantum (super)algebras.
  {C}onnection of {D}rinfeld's and {C}artan-{W}eyl realizations for quantum
  affine algebras}, \href{http://arxiv.org/abs/hep-th/9404036}{{\tt
  arXiv:hep-th/9404036}}.

\bibitem{ChaPre91}
\curlanguage{English}
V.~Chari and A.~Pressley, \emph{Quantum affine algebras},
  \href{http://dx.doi.org/10.1007/BF02102063}{Commun. Math. Phys.} \textbf{142}
  (1991), 261--283.

\bibitem{KliSch97}
\curlanguage{English}
A.~Klimyk and K.~Schm{\"u}dgen, \emph{Quantum groups and their
  representations}, Texts and Monographs in Physics, Springer, Heidelberg,
  1997.

\bibitem{Jim86a}
\curlanguage{English}
M.~Jimbo, \emph{A $q$-analogue of {$\mathrm U(\mathfrak{gl}(N + 1))$}, {H}ecke
  algebra, and the {Y}ang--{B}axter equation},
  \href{http://dx.doi.org/10.1007/BF00400222}{Lett. Math. Phys.} \textbf{11}
  (1986), 247--252.

\end{thebibliography}
\end{document}